\documentclass[sigconf]{acmart}

\AtBeginDocument{%
  \providecommand\BibTeX{{%
    \normalfont B\kern-0.5em{\scshape i\kern-0.25em b}\kern-0.8em\TeX}}}

\setcopyright{acmcopyright}
\copyrightyear{2023}
\acmYear{2023}
\setcopyright{rightsretained}
\acmConference[FAccT '23]{2023 ACM Conference on Fairness, Accountability, and Transparency}{June 12--15, 2023}{Chicago, IL, USA}
\acmBooktitle{2023 ACM Conference on Fairness, Accountability, and Transparency (FAccT '23), June 12--15, 2023, Chicago, IL, USA}
\acmDOI{10.1145/3593013.3594037}
\acmISBN{979-8-4007-0192-4/23/06}

\usepackage[normalem]{ulem}
\usepackage{scalerel}
\usepackage{multirow}
\usepackage{enumitem}
\usepackage{listings}
\usepackage{xparse}
\usepackage[utf8]{inputenc}
\usepackage{xcolor}
\usepackage{soul}
\usepackage{colortbl}
\usepackage{makecell}
\usepackage{hyperref}
\usepackage[nameinlink]{cleveref}
\usepackage{array}
\usepackage{comment}

\begin{document}

%%
%% The "title" command has an optional parameter,
%% allowing the author to define a "short title" to be used in page headers.
\title[Cross-functional Collaboration in Industry AI Fairness]{Investigating Practices and Opportunities for Cross-functional Collaboration around AI Fairness in Industry Practice}

%%% Title options %%%
% Uncovering the Under-recognized Bridging and Piggybacking Work in Cross-functional Collaboration for AI Fairness in Industry Practice}

% Uncovering the Practices and Opportunities for Cross-functional Collaboration for AI Fairness in Industry Practice

% Opportunities to Support Cross-functional Collaboration for AI Fairness in Industry Practice 

%%
%% The "author" command and its associated commands are used to define
%% the authors and their affiliations.
%% Of note is the shared affiliation of the first two authors, and the
%% "authornote" and "authornotemark" commands
%% used to denote shared contribution to the research.

\author{Wesley Hanwen Deng}
\email{hanwend@cs.cmu.edu}
\affiliation{%
  \institution{Carnegie Mellon University}
  \streetaddress{5000 Forbes Ave}
  \city{Pittsburgh}
  \state{PA}
  \postcode{15213}
  \country{USA}
}

\author{Nur Yildirim}
\email{yildirim@cmu.edu}
\affiliation{%
  \institution{Carnegie Mellon University}
  \streetaddress{5000 Forbes Ave}
  \city{Pittsburgh}
  \state{PA}
  \postcode{15213}
  \country{USA}
}

\author{Monica Chang}
\email{monicach@andrew.cmu.edu}
\affiliation{%
  \institution{Carnegie Mellon University}
  \streetaddress{5000 Forbes Ave}
  \city{Pittsburgh}
  \state{PA}
  \postcode{15213}
  \country{USA}
}

\author{Motahhare Eslami}
\email{meslami@cs.cmu.edu}
\affiliation{%
  \institution{Carnegie Mellon University}
  \streetaddress{5000 Forbes Ave}
  \city{Pittsburgh}
  \state{PA}
  \postcode{15213}
  \country{USA}
}

\author{Kenneth Holstein}
\email{kjholste@cs.cmu.edu}
\affiliation{%
  \institution{Carnegie Mellon University}
  \streetaddress{5000 Forbes Ave}
  \city{Pittsburgh}
  \state{PA}
  \postcode{15213}
  \country{USA}
}

\author{Michael Madaio}
\authornote{Michael began this work while at Microsoft Research.}
\email{madaiom@google.com}
\affiliation{%
  \institution{Google Research}
  \city{New York}
  \state{New York}
  \postcode{15213}
  \country{USA}
}

%%
%% The abstract is a short summary of the work to be presented in the
%% article.
\begin{abstract}
An emerging body of research indicates that ineffective cross-functional collaboration -- the interdisciplinary work done by industry practitioners across roles -- represents a major barrier to addressing issues of fairness in AI design and development. In this research, we sought to better understand practitioners' current practices and tactics to enact cross-functional collaboration for AI fairness, in order to identify opportunities to support more effective collaboration. We conducted a series of interviews and design workshops with 23 industry practitioners spanning various roles from 17 companies. We found that practitioners engaged in \textit{bridging} work to overcome frictions in understanding, contextualization, and evaluation around AI fairness across roles. In addition, in organizational contexts with a lack of resources and incentives for fairness work, practitioners often \textit{piggybacked} on existing requirements (e.g., for privacy assessments) and AI development norms (e.g., the use of quantitative evaluation metrics), although they worry that these tactics may be fundamentally compromised. Finally, we draw attention to the \textit{invisible labor} that practitioners take on as part of this bridging and piggybacking work to enact interdisciplinary collaboration for fairness. We close by discussing opportunities for both FAccT researchers and AI practitioners to better support cross-functional collaboration for fairness in the design and development of AI systems.\looseness=-1

%Overall, we found that practitioners on cross-functional AI teams performed various forms of invisible labor to make fairness work possible in practice. % we argue that such labor has been similarly under-supported by FAccT research to date. %Building on an understanding of the bridging work and piggybacking work that practitioners perform, along with the underlying challenges practitioners remain facing, 
% initiatives (e.g., around privacy), business logic, and established AI development process (e.g., preference of quantification as evaluation)
%fairness work in industry practice and beyond. 
\end{abstract}

%%
%% The code below is generated by the tool at http://dl.acm.org/ccs.cfm.
%% Please copy and paste the code instead of the example below.
%%

%%
%% Keywords. The author(s) should pick words that accurately describe
%% the work being presented. Separate the keywords with commas.

%% A "teaser" image appears between the author and affiliation
%% information and the body of the document, and typically spans the
%% page.
%\begin{teaserfigure}
  %\includegraphics[width=\textwidth]{figures/teaser.pdf}
  %\caption{ 
%}
  %\Description{Two images of Zeno, the API and the UI.}
  %\label{fig:teaser}
%\end{teaserfigure}

%%
%% This command processes the author and affiliation and title
%% information and builds the first part of the formatted document.
\maketitle

\section{Introduction}

Addressing unfairness in AI systems is a fundamentally socio-technical challenge, requiring the integration of skills and expertise across many different areas, including approaches from the social sciences to understand what (un)fairness means for particular communities and sociocultural contexts, approaches from user research and design to understand (un)fairness in particular use cases and domains, as well as technical skills to address unfairness in AI system design \cite{green2021contestation,forsythe2001studying,Gebru2022Hierarchy}.
Yet a growing body of literature suggests that such integration of skills and expertise that comes in the form of  
\textit{cross-functional collaboration} \cite{Henke1993CrossFunctionalTG, song1997cross, kahn1996interdepartmental} --- a term used in industry settings to describe collaboration among diverse roles with various disciplinary backgrounds \cite[cf.][]{rakova2021responsible} --- is often absent or ineffective in industry AI fairness work \cite{almahmoud2021teams, deng2022exploring, nahar2021collaboration, zhang2020data, yildirim2022experienced, passi2019problem, passi2018trust, madaio2022assessing}. Prior work has identified several challenges around cross-functional collaboration for AI fairness, such as differing awareness of AI fairness across roles \cite{nahar2021collaboration, almahmoud2021teams, zhang2020data}, mismatched expectations for measuring fairness \cite{madaio2022assessing, deng2023understanding, passi2018trust}, and an absence of tools to support collaborative work across roles \cite{almahmoud2021teams,deng2022exploring,wong2022seeing}. These challenges have been shown to hinder teams in effectively addressing fairness issues \cite{madaio2022assessing, deng2023understanding, nahar2021collaboration, almahmoud2021teams, zhang2020data}. However, little is known regarding (1) whether and how industry practitioners navigate such challenges to overcome collaboration barriers, and (2) what opportunities exist to improve cross-functional collaboration around fairness in AI.   

To investigate this, we conducted a two-stage study with 23 industry AI practitioners spanning 17 companies and various roles (e.g., data scientists, UX practitioners, product managers, and subject matter experts), who have previously worked with other roles in their company to tackle fairness-related issues. In our study, we first conducted semi-structured interviews to understand practitioners’ current practices and challenges around cross-functional collaboration. We then conducted workshops to bring together participants across multiple roles and companies, to better understand common practices and envision future opportunities for creating more effective cross-collaboration in AI fairness. 

We found that practitioners go beyond their current job descriptions to undertake a range of ``bridging’’ roles, aimed at fostering shared understandings around AI fairness, translating and contextualizing abstract fairness concepts for other roles, and aligning expectations around fairness evaluation across different roles (Section \ref{bridging}). In order to effectively collaborate in organizations constrained by a lack of resources or incentives for AI fairness work \cite[cf.][]{rakova2021responsible,madaio2020co}, practitioners also adopted ``piggybacking’’\footnote{Informed by prior research on environmental sustainability and social justice work in industry settings \cite[cf.][]{Brown1998MakingBI, burnie2016piggybacking, robinson2010save}, we use ``piggybacking'' to refer to the process of \emph{``identifying potential allies with similar or overlapping interests, and utilize and `piggyback' on existing organizational resources and programs as much as possible''} \cite{Brown1998MakingBI}.} tactics to facilitate collaboration around AI fairness, although they worried the use of such tactics might compromise their values in the long term (Section \ref{piggybacking}). Furthermore, participants shared frustration around the often invisible labor involved in supporting cross-functional collaborations (Section \ref{invisible_labor}).

Building on an understanding of these bridging and piggybacking practices and the underlying collaboration challenges they are intended to address, we discuss opportunities for both researchers and industry AI teams to better support cross-functional collaboration for AI fairness work in industry practice (Section \ref{Discussion}). In particular, we discuss tools and processes that could support ``bridging'' work, opportunities and risks around "piggybacking" tactics for carrying out AI fairness collaborations, and we discuss ways to make invisible labor in collaborations around AI fairness more visible to (and ideally valued by) teams and organizations. 

Overall, this paper contributes an in-depth understanding of industry practitioners' current practices to facilitate cross-functional collaboration across roles and organizations, identifying ``bridging'' work and ''piggybacking'' as two major approaches. These practices serve as a starting point for practitioners to navigate cross-functional collaboration challenges in AI fairness work. In addition, we identify implications for FAccT researchers, practitioners, and organizations to better support cross-functional collaboration in AI fairness work.

\section{Background and Related Work}
%\subsection{Understanding and supporting AI fairness practices in industry contexts}
%\michael{So, it's not actually clear to me that we need a standalone subsection for this - it seems like everything substantive we have to say in 2.1 (lines 90-98) is actually specifically about cross-functional collaboration for fairness, which is 2.2.}
%Responsible AI processes, guidelines, and toolkits have been developed to incorporate fairness as a core part of AI design, development, and deployment \cite{MSRchecklist, googlePAIR, AIF360API, AIX360API, FairlearnAPI, bird2020fairlearn, EYTrust}. In parallel, 

A growing body of research has empirically investigated industry practitioners' current practices and challenges in addressing issues of fairness \textit{in practice}, during the design and development of AI systems \cite[e.g.,][]{cramer2018assessing, passi2018trust, passi2019problem, holstein2019improving, madaio2020co, Kaur2020InterpretingIU, rakova2021responsible, madaio2022assessing, lee2021landscape, richardson2021towards, deng2022exploring, deng2023understanding, wang2023designing, smith2022real}. Among other findings, this work has suggested that effectively tackling socio-technical challenges like AI fairness (and more broadly, building more responsible AI) requires substantial interdisciplinary collaboration among multiple roles \cite{passi2019problem, holstein2019improving, madaio2020co, rakova2021responsible, deng2022exploring}. For example, based on fieldwork with a corporate data science team, Passi and Jackson highlighted that in order to build more responsible and trustworthy AI systems, data scientists engage and negotiate with business and product teams throughout the AI development lifecycle  \cite{passi2018trust}. Rakova et al. suggested that addressing AI fairness issues requires AI developers to better understand the needs of stakeholders from different backgrounds (e.g., domain experts) \cite{rakova2021responsible}. 

%In this work, we use \textit{``cross-functional collaboration,''} a term commonly used in industry settings to describe the interdisciplinary work done by industry practitioners across roles \cite[cf.][]{Henke1993CrossFunctionalTG, song1997cross, kahn1996interdepartmental, rakova2021responsible}. 
 %In the following section, we synthesize the challenges around collaboration in AI fairness.

%\subsection{Challenges in Cross-functional Collaboration for AI Fairness in Industry Practice}
%\footnote{In Section \ref{Findings}, we use these acronyms to refer to participants' roles, in the format of [role acronym]$_$[assigned number] (e.g., DS$_$5)} \cite{amershi2019software, zhang2020data}. %However, as mentioned briefly in the last section, challenges exhibited in the cross-functional collaboration in AI fairness.
%In industry AI teams, designing, developing, and monitoring AI products and services requires effective communication and collaboration among a diverse range of roles with different disciplinary backgrounds — including software engineers (SWE), data scientists (DS), machine learning (ML) engineers, user experience (UX) designers, product managers (PM), and subject matter, or domain experts (SMEs). \dhw{Try to think about how to merge this to intro (as a foreshadowing of the roles that we are discussing in this paper.)}

However, despite its importance, prior research suggests cross-functional collaboration is often absent or ineffective in industry AI fairness work. AI fairness work can accentuate challenges for cross-functional collaboration that are present in other areas of AI development, and may introduce new challenges \cite[cf.][]{yang2020re, piorkowski2021ai, yang2020re, subramonyam2022solving, kross2021orienting, nahar2021collaboration, zhang2020data, amershi2019software, yildirim2022experienced}. For instance, given the fundamentally socio-technical and contested nature of AI fairness, the metrics and methods used to conceptualize, evaluate, and address AI fairness issues can vary substantially across disciplines \cite{Gebru2022Hierarchy, selbst2019fairness, yurrita2023disentangling}. In practice, these properties lend themselves to communication breakdowns and ineffective collaboration around AI fairness \cite{passi2018trust, passi2019problem, madaio2022assessing, deng2022exploring, Hutchinson2022EvaluationGI}. Passi and Barocas found that misalignments around problem formulation between data scientists and business teams can contribute to fundamental fairness issues from the early problem formulation phases of a project \cite{passi2019problem}. Madaio et al. found that practitioners across roles often defaulted to using their existing performance metrics for assessing the aggregate performance of their AI systems, when those metrics may not be best suited to identifying disparities in models' performance for \textit{dis}aggregated subgroups for particular use cases or contexts \cite{madaio2022assessing}. Currently, we lack processes for practitioners across disciplines to decide on appropriate metrics to assess disparities in model performance. These challenges compound when collective decision-making across multiple areas of expertise is required. \looseness=-1

%This removed opportunities to 
%For instance, in a study of industry collaboration challenges in building AI systems, Nahar et al. [2022] surfaced the lack of awareness around fairness between software engineers and data scientists \cite{nahar2021collaboration}. In other empirical work, Madaio et al. [2022] found that mismatched expectations for model performance metrics from different roles posed obstacles to collectively conducting fairness assessments %(i.e., identifying performance disparities for different demographic groups) \cite{madaio2022assessing}. In their work studying how AI practitioners across roles communicate the quality of AI systems, Almahmoud et al. [2021] found that cross-functional AI teams often neglected fairness as part of the model qualities; product managers in the study reported a lack of tools and process to support discussing fairness as part of the AI system qualities with data scientists \cite{almahmoud2021teams}.

Relatedly, although recent research has begun to develop tools and processes for cross-functional collaboration in AI development \cite[e.g.,][]{lam2023model, cabrera2022did, park2021facilitating, subramonyam2022solving, moore2023failurenotes}, existing tools and processes designed for tackling AI fairness issues \cite[e.g.,][]{AIF360API, FairlearnAPI, EYTrust, cabrera2019fairvis, cabrera2021discovering} are largely designed to be used by technical roles working in isolation, and are not designed to support collaboration across roles \cite{deng2022exploring, almahmoud2021teams, lee2021landscape, wong2022seeing, zhang2020data}. For example, in their work studying how AI practitioners use fairness toolkits, Deng et al. identified that data scientists lack efficient tools and processes to translate and incorporate domain experts' knowledge into fairness analyses \cite{deng2022exploring}. When studying how teams communicate about and evaluate the quality of ML models, Almahmoud et al. found that practitioners felt current tools and resources made it challenging to have cross-functional conversations around fairness, as these tools are often tailored to evaluate and report model qualities like accuracy rather than broader sociotechnical concepts such as fairness \cite{almahmoud2021teams}, potentially contributing to the reification of technical values in AI \cite{Gebru2022Hierarchy,birhane2022values}. \looseness=-1

Finally, despite growing effort toward engaging diverse roles in AI development \cite{muller2019data, subramonyam2021towards, yang2020re, park2021facilitating, mao2019data, liao2023designerly}, prior literature has found that not all relevant roles in AI teams are incentivized or invited to collaborate around AI fairness \cite{nahar2021collaboration, zhang2020data}. For example, Zhang et al. found that industry AI teams often treated fairness work primarily as a technical matter; as a consequence, roles with the most relevant domain, legal, policy, or lived expertise were often left out of the process \cite{zhang2020data}. While prior work has discussed challenges to collaboration around fairness, in the current work we seek to understand what industry AI practitioners currently do to support and enable cross-functional collaboration for AI fairness, with the goal of identifying opportunities to better support these efforts. %Thus, in this study, we take the initial step to explore the practices, tactics, and opportunities for cross-functional collaboration around AI fairness in industry practice.

%\dhw{I think alternatively we could also move RQs here. The current wording is indeed quite repetitive and I'm wondering how to make it better.}

\section{Methods} \label{Methods}

To investigate our research questions, we conducted a two-stage study with 23 industry practitioners across 7 roles, involving semi-structured interviews followed by group workshops. We first conducted semi-structured interviews to understand participants’ current practices and challenges around cross-functional collaboration in AI fairness work. In the next stage, we invited participants spanning different roles and organizations to workshops to collaboratively explore opportunities to better support cross-functional collaboration around AI fairness. 
%We include all the interview and workshop study protocols in the supplementary materials.

\subsection{Participants}
We adopted a purposive sampling approach \cite{campbell2020purposive}, with the aim of recruiting industry AI practitioners from diverse roles (e.g., data scientists, UX researchers, product managers, etc) who have worked on addressing AI fairness in their role. During the study, all participants self-reported that they had experience collaborating with other roles in their work on AI fairness. Table \ref{participant_table} provides an overview of participants' product areas and roles. Throughout the paper, we use [Acronym of the role][Role number] to refer to our participants and we use [W][Workshop session number] to refer to the workshop session participants attended.

We recruited our participants through direct contacts at large technology companies, through recruitment posts on social media (e.g., Linkedin and Twitter), and snowball sampling from those participants. In total, 25 practitioners completed the recruitment screening form for our interview, of whom 18 met our recruitment criteria, responded to our interview study invitation, and participated in the study. In the end, 17 participants completed the interview study. We invited all interview participants to attend a workshop session. 6 out of 17 interview participants responded to our invitation and joined the workshops (W1 [DS2, DS4, UX5], W2 [DS1, MLE1, DS5]). We then sent another round of workshop recruitment to new participants, using a purposive sampling approach similar to the previous round (e.g., direct contacts and social media) and received 42 responses. In the end, 19 of those responded to our scheduling email, with 6 able to participate in the second round of workshops (W3 [PM3, UX6], W4 [SME1, DS6], W5 [DS1, SWE2, SME2])\footnote{DS1 volunteered to join another round of the workshop. We believed that engaging DS1 with new participants from other organizations could lead to valuable insights.}. We scheduled each workshop to include participants spanning different roles and organizations. Similar to prior work studying AI fairness with industry practitioners \cite{lee2021landscape, deng2022exploring, richardson2021towards}, we encountered a large drop-out rate, potentially due to the sensitivity of the topic of AI fairness. In addition, scheduling the workshops with participants from different roles and companies was constrained by practitioners' joint availability, further adding challenges to participation in the study. 

% \dhw{We also need to explain somewhere why, for example, W2 only had technical roles (DS1, MLE1, and DS5) -- mainly due to the low retention of interview participants who replied to the workshop invite and the constraints around scheduling from practitioners. (The invisible labor of conducting user study!!!!)}

All participants were compensated at a rate of \$35 per hour for their participation. In addition, for both interviews and workshops, we told participants that we would not ask them to reveal any confidential or personally identifying information about their colleagues and that we would anonymize all responses at the individual, team, and organization levels. Finally, participants were told that they were free to skip any questions they were uncomfortable answering, and were free to leave the workshop session at any time for any reason. This study was approved by our institution's IRB.

% \dhw{TODO: how do we communicate the participants group?}\textbf{17 participants participated in the interview study}: DS1 - DS5; UX1 - UX5; PM1 - PM2; MLE1 - MLE2; SWE1; RS1 - RS2. \textbf{12 participants who participated in the five workshops (W1 - W5)}: \textbf{W1}: [DS2, DS4, UX5]. \textbf{W2}: [DS1, MLE1, DS5]. \textbf{W3}: [PM3, UX6]. \textbf{W4}: [SME1, DS6]. \textbf{W5}: [DS1, SWE2, SME2]

%\begin{comment}
\begin{table*} 
  \begin{tabular}{ c c c c c} 
     \toprule
     \textbf{Role or Job Titles} & \textbf{Company Employee Number} & \textbf{Gender} & \textbf{Location} \\ 
     \midrule
     Data Scientist (DS) (6) &  25,000 and more (9) & Female (7) & US (15) \\ 
     UX Practitioners (UX) (6) &  5,000 - 24,999 (2) & Male (15) & India (2)  \\ 
     Product Manager (PM) (3) &  1,000 - 4,999 (3) & Non-binary (1) & Sweden (2) \\ 
     ML Engineer (MLE) (2) &  250 - 999 (0) &  & Australia (1) \\ 
     Software Engineer (SWE) (2) &  50 - 249 (1) &  & France (1) \\
     Research Scientist (RS) (2) &  10 - 49 (2) &  & Netherlands (1) \\
     Subject Matter Expert (SME) (2) &  &  & Mexico (1) \\

     \bottomrule
    \end{tabular}
    \caption{Overall demographics and background of our 23 study participants. Next to each demographic information, we include the number of the participants within that demographic group in parenthesis.}
    \label{participant_table}
\end{table*}

%\textbf{17 participants participated in the interview study}: DS1 - DS5; UX1 - UX5; PM1 - PM2; MLE1 - MLE2; SWE1; RS1 - RS2. \textbf{12 participants who participated in the five workshops (W1 - W5)}: \textbf{W1}: [DS2, DS4, UX5]. \textbf{W2}: [DS1, MLE1, DS5]. \textbf{W3}: [PM3, UX6]. \textbf{W4}: [SME1, DS6]. \textbf{W5}: [DS1, SWE2, SME2]

\subsection{Study Design}
\subsubsection{Stage one: semi-structured interviews} %Understanding Current Practices and Challenges around Cross-Functional Collaboration for AI Fairness}

To understand practitioners’ current practices around cross-functional collaboration for fairness in AI, we conducted 17 individual semi-structured interviews with practitioners from diverse roles from 12 organizations, all of whom had some experience working on AI fairness. We adopted a directed storytelling approach \cite{evenson2006directed,gray2010gamestorming} in the interviews, each of which lasted roughly an hour. We first asked participants to describe their current practices around cross-functional collaboration in AI fairness, with a specific focus on artifacts, tools or resources their team used as part of that work. For example, we asked participants \emph{``What tools have you been using to collaborate with other roles on your team around AI fairness?''} We probed deeper into challenges participants had encountered by asking follow-up questions like \emph{``Were there disagreements or tensions between people from different disciplinary backgrounds?’’} As participants shared specific collaboration challenges they had encountered, we invited them to share specific activities or strategies they adopted to address those, by asking questions like \textit{``How did your team attempt to tackle these challenges?’’} and \textit{``How effective were your team’s approaches?’’}\looseness=-1
 
\subsubsection{Stage two: workshops} %Exploring the Opportunities to Navigate Cross-Functional Collaboration Challenges for AI Fairness  }

In the second stage, we conducted an iterative series of five workshops to create a space for multiple industry practitioners from \textit{different roles, teams, or organizations} to share their experiences, discuss how their experiences related to other participants, and identify opportunities for improving cross-functional collaboration for AI fairness. 12 participants attended the workshops, including six participants from the interviews and six new participants. For the first two workshops (W1, W2), we recruited participants from the interviews in stage one and asked them to discuss preliminary findings from the interview study, to validate these preliminary results (or discuss tensions or differences), and share additional context from their own experiences. We then had participants conduct a speed boat activity \cite{pavelin2014ten}---a common workshop activity used in design workshops for facilitating discussion among multiple stakeholder groups. In the activity, we shared challenges to collaboration on AI fairness that we identified in our preliminary findings, and asked participants to collectively select one such challenge and imagine themselves to ``be on the same boat'' with each other to address the challenge. Drawing on the metaphor of a boat at sea, participants identified their goals, tailwinds and headwinds (e.g., enabling or hindering factors), and other aspects of their team or organization that might impact their ability to effectively collaborate on fairness with team members from other disciplines.

After running the first two workshop sessions, we observed that our participants particularly needed to explore the opportunities they saw for ideal cross-functional collaboration. We thus revised our workshop protocol to provide a space for brainstorming desired cross-collaboration opportunities via ``journey mapping'' \cite{lee2022hci}---a technique from user experience research that can be used to identify interactions between different stakeholders over time and support participants in moving from the current state to an ideal state. Before joining the workshops, participants watched a short tutorial video we prepared and filled out a journey map defining their teams' phases of AI development and the stakeholders involved in each phase, based on the nature of their team and organization. For each phase they listed, we asked participants to share more details around the work they did in this phase (related to fairness specifically); the other roles and stakeholders that are currently involved (and the roles they wished were more involved); the artifacts or resources they currently used (and that they wished to have); as well as what worked well and the pain points of each phase. 

We compiled all participants’ journey maps to a Miro board before each workshop session. During the workshop, we spent the first thirty minutes having each participant quickly present their journey maps so that they could learn more about each others’ processes. We then used another thirty minutes to have the participants discuss and co-construct an ``ideal’’ journey map using a journey map template we offered, focusing on envisioning the opportunities to better navigate cross-functional collaboration. Note that, similar to prior HCI research using journey maps in their study design (e.g., \cite{lee2022hci}), our goal for the workshop was not to produce a perfect journey map as an artifact-based contribution of the research, but instead, we used the process of discussing and producing journey maps to elicit participants’ perceived challenges, needs, and opportunities, as well as to scaffold the workshop discussions with specific details from participants' roles, teams, and organizations.\looseness=-1

\subsection{Data Analysis}

Our study sessions yielded approximately 23 hours of audio that we transcribed. To analyze our interview and workshop transcripts, we used inductive thematic analysis, a common qualitative data analysis method in HCI \cite{Braun2006-ts}. Six of the authors conducted an open coding of a subset of the transcripts, then discussed their codes with the entire research team. After the team reached consensus on the format and granularity of the codes, two authors independently coded all transcripts and reconvened to resolve any major disagreements through discussion. For example, codes include \emph{``MLE1 decided to use their spare time to develop shareable documentations and materials to help other team members learn more about AI fairness.''} Then, four of the authors iteratively grouped the codes to identify higher-level themes. For example, one lower-level theme around practitioners' current practices was \emph{``Participants leverage numerical, measurable metrics to translate the impact of fairness to AI developer teams''}, which was later grouped into the higher-level theme of \emph{``Piggybacking on the quantification culture of AI development.''} We present key themes from our data in the following Findings sections.\looseness=-1

\section{Findings} \label{Findings}

\subsection{Bridging Gaps in Understanding and Evaluation to Improve Collaboration} \label{bridging}

 We found that participants (spanning a range of formal roles) took on critical bridging roles by identifying and creating opportunities to foster their team's learning about AI fairness, contextualizing abstract concepts and guidelines to make AI fairness work concrete, and aligning mismatched goals and metrics for fairness evaluations. Throughout this section, we highlight how participants use these bridging activities to attempt to overcome barriers to cross-functional collaboration around AI fairness. 

\subsubsection{Bridging the gaps in incompatible disciplinary evaluations around AI fairness}\label{evaluation}

%Alignment from UX PM SWE DS, mistranslation happens, missing the opportunities to address AI fairness issues from a socio-technical lens. 
%Prior work suggested that collaboration challenges arise around mismatched goals between different disciplinary roles when evaluating the quality of ML systems \cite{nahar2021collaboration}. 
We find that our participants take on bridging work to overcome tensions in the methods or metrics that various disciplines use to assess or measure fairness in AI systems, metrics which may be incompatible with each other. For example, participants in our study reported that technical roles on their teams (DS, MLE) tend to  evaluate their models by \emph{``mainly focusing on the output of the models they built without thinking about how [these models] interact with real customers’’} (SWE2). In contrast, user- and product- facing roles (e.g., UX designers and PMs) often have a better understanding of \emph{``shareholders and customers' concerns''} (PM2) but may lack an understanding of how to translate their understanding into effective evaluations of fairness (e.g., in ways that respond to customers' concerns).\footnote{See \citet{madaio2022assessing} for a discussion of the risks of prioritizing shareholders, customers, or other business-oriented stakeholder groups over marginalized communities who may be most impacted by algorithmic systems.} As a consequence, multiple participants mentioned that in order to facilitate communication about fairness among team members with diverse backgrounds, they needed to bridge the goals for fairness assessments between technical roles focused primarily on the model outputs (i.e., model-focused evaluation) and user- and product- facing roles who tend to focus on biases and harms perceived by users (i.e., user-focused evaluation). %\dhw{Here I spent more time on explaining what I meant by "model-focused evaluation" vs. "user-focused evaluation"}

Participants shared that they bridged these evaluation gaps by initiating and organizing meetings for cross-functional teams to align model-focused and user-focused fairness evaluations. For example, PM1 repurposed some of their regular team-level all-hands meetings, which were originally designed for team members to report on their work progress and goals, to \emph{``co-evaluation meetings''} (PM1) for fairness issues. In particular, PM1 spent extra effort designing activities to scaffold technical roles and user-facing roles in understanding and aligning each others' perspectives on evaluating fairness issues:
%\begin{quote}
    \emph{``I will have the entire team together and have the failure mode effect analysis, we see what’s happening with the false positive rate and false negative rate of the model for our use cases and making sure that we always align on this [...] for every single step [in building the model], our team makes sure that there is a fairness requirement from the product team and there is a specific guideline of implementation from the engineers.''} (PM1)
%\end{quote}

UX5 shared that they organized similar \emph{``co-evaluation sessions’’} (UX5) in their company, with the purpose to create spaces to understand the differences and similarities between the evaluation metrics for AI fairness that different roles employed.
%\begin{quote}
    \emph{``In these sessions, I come up with hierarchies and frameworks whilst everybody else was talking talking talking [...] and sharing these artifacts in the moment [...] and showing what are the connections between different evaluations people were just talking about [...] and then people made the connections between the AI [models] and the product and they suddenly started to collaborate because they finally understood each other's goals [...] and started to use similar terminologies.''} (UX5)\looseness=-1
%\end{quote}

% \dhw{potentially: one sentence to expand on the previous quotes. Re: Michael and Shaun's review}
However, participants shared that, while they attempted to bridge the gaps between model-facing and user-facing evaluations of
%quantitative and qualitative evaluation 
AI fairness, the culture of AI development (broadly speaking) often prioritizes quantitative over qualitative evaluation approaches \cite{Gebru2022Hierarchy, birhane2022values}, making this bridging work around fairness evaluation less effective. In Section \ref{quantification}, we expand on practitioners’ strategies on navigating the mismatches between quantitative and qualitative evaluation approaches in organizations that disincentivize fairness work.

\subsubsection{Creating collaborative processes and spaces to bridge gaps in understanding about AI fairness} \label{understanding}

We found that the incompatibility between different teams was not limited to fairness evaluation methods and metrics. In fact, most participants shared that there were crucial differences in their understanding of AI fairness in general, which introduced challenges for effective collaboration. Without a common grounding of what AI fairness entails for the specific domain they were working in, \emph{``disagreements in terms of definitions of bias and fairness will affect how people use the [fairness] toolkits and all the downstream collaborations’’} (RS1). %\footnote{In the Discussion, we expand on the plethora of largely incompatible fairness metrics, and so these disagreements have very real implications for how fairness is measured}
During the workshop (W1), UX5 told us that the conversations around fairness among their team members \emph{``stayed at a superficial level and went nowhere''} when team members failed to align their own understandings of what ``fairness'' means with what it means to their users --- in the context of their specific application. In another workshop (W5), SME2 shared in their ``journey map” activity that they realized some of their coworkers were \emph{``not aware of representational harms\footnote{Representational harms are fairness-related harms that involve how groups of people are represented by algorithmic systems, including stereotyping, demeaning, or erasure of particular groups entirely \cite[e.g.,][]{crawford2017trouble,blodgett2020language,wang2022measuring}} caused by AI systems at all.’’} This made them realize the importance of \emph{``check[ing] each others’ understanding of [fairness] problems before just diving into the conversation around fixing the problems.''}\looseness=-1

To bridge these gaps and inconsistencies in AI practitioners’ understanding of AI fairness, participants designed various collaborative activities that were intended to scaffold conversations among cross-functional team members. These activities range from small team design workshops to company-wide hackathons. For instance, as a UX designer working on image captioning and product recommendation applications, UX1 held design workshops \emph{``similar to those workshops us designers often conduct with external clients and users”} with their team members working on AI fairness. In these workshops, UX1 led the conversations with their team members working on different aspects of the products to explore what \emph{``a fair product recommendation system mean[s]''} (UX1).

However, participants mentioned that team members, especially those who were new to the topic of AI fairness, often struggled with what and how to discuss. To overcome these communication barriers, participants drew on toolkits to scaffold the design process. For example, UX4 incorporated a toolkit they often used in user research sessions called \emph{``ideation cards’’}---a deck of cards including a hundred questions concerning the value and ethics around product design---to facilitate cross-role design workshops and better engage team members in asking questions probing how their AI products might cause fairness issues under different scenarios. UX4 shared that ideation cards helped their team to have constructive debates around the meaning of being fair to different stakeholders who might be affected by the AI service.\looseness=-1 

To foster participation and engagement in conversations around AI fairness, some participants described how they tailored collaborative activities to the skills and knowledge of specific roles. For example, UX2, PM2, and UX5 mentioned hosting company-wide hackathons, a familiar and engaging format for technical roles like engineers, to better engage engineers in critically examining their understanding of AI fairness when building AI applications . These hackathons often happened \emph{``at the problem formulation stage when the team starts to work on algorithmic fairness or transparency relevant topics’’} (UX2), serving as a chance for team members to begin the conversation around fairness issues of their AI products and learn more about each others’ perspectives on fairness. During W1, while discussing the "tailwinds" of their current collaboration, UX5 brought up how the hackathon event they designed for building more responsible AI helped engage diverse roles in conversation about AI fairness. During this workshop session, other participants in technical roles (DS2 and DS4) expressed their interest in participating in similar events and implementing this format in their own organizations to potentially promote collaboration on AI fairness initiatives across roles.

\subsubsection{Developing educational resources and documentation to support understanding about fairness}
\label{education}

Complementing the efforts for building a common ground between team members about AI fairness, participants also reported developing documentation and othe resources (in addition to hosting workshops) that aimed to increase their team's knowledge about AI fairness and facilitate conversations about this topic in collaborations. In particular, participants with strong technical backgrounds (DS1, MLE1, RS2, DS6) who report being familiar with the state of the art of technical fair AI research literature shared that they created accessible educational materials to help their team members learn about technical AI fairness concepts (e.g., fairness metrics, bias mitigation algorithms, and their limitations). For example, DS6 created a guidebook covering \emph{``common fairness metrics, rationales for some bias mitigation algorithms, and also different types of harms that could be caused by algorithms.’’} 

MLE1 worked in a small start-up company offering consulting services for building responsible AI and self-reported being the most knowledgeable team member around AI fairness in their growing engineering team. After repeatedly getting similar questions from colleagues around \emph{``AI fairness concepts and  confusions when reading some paper they saw from FAccT,’’} MLE1 decided to use their spare time\footnote{In section~\ref{invisible_labor}, we discuss the consequences of AI team members needing to use their spare time to develop resources to bridge disciplinary gaps.} to develop \emph{``shareable documentations and materials''} to provide their team members with a \emph{``free crash course[s]''} that explained basic AI fairness concepts using accessible language. This documentation then became standard onboarding materials for new employees in their company to learn more about fairness in AI.\looseness=-1 

\subsubsection{Translating and contextualizing abstract AI fairness concepts in concrete terms} \label{contextualization}

Participants in our study repeatedly mentioned that publicly available AI fairness guidelines and tutorials (and the AI fairness concepts presented in them) are \emph{``usually too abstract’’} (DS4) for tackling the AI fairness issues in their organizations. As a result, when attempting to follow these AI fairness guidelines and tutorials, technical roles often found them \emph{``not always relevant to the task at hand’’} (MLE2) and struggled to \emph{``understand which fairness metrics or techniques to use for the specific application''} (DS6). To this end, we observed translation and contextualization work between technical roles and user-facing roles to better understand what fairness means for their specific domain and use case or user populations.\looseness=-1

Participants in technical roles (DS, SWE) --- who are often responsible for directly conducting algorithmic fairness analyses --- often proactively initiated collaboration with user-facing roles (PM, customer services) to better contextualize abstract fairness metrics used in analysis in real world scenarios. For instance, as a data scientist, DS5 worked closely with UX researchers to \emph{``build a glossary to contextualize the abstract concept of fairness metrics''} such as equalized odds and demographic parity using real-world scenarios in their AI services for healthcare. Similarly, SWE1 reported that they went beyond assessing model fairness through fairness metrics, trying to contextualize the analysis through conversations with customer service managers. Participants shared that the contextualizing process sensitized data scientists about \emph{``how their model might interact with users in an unintended, harmful way''} (DS5), helping technical roles learn more about other roles' perspectives to better inform their fairness analysis.\looseness=-1

Furthermore, we found that multiple user- and product-facing roles (UX, PM) often spent extra effort contextualizing abstract AI fairness guidelines to their actual practices for other team members, in order to \emph{``get the entire teams on the same page around how the models might cause fairness issues when they are interacting with users''} (PM1). In particular, PM1 annotated the AI fairness guidelines developed by their company (a technology company with over 25,000 employees) with concrete examples and prompts to explore, for instance, \emph{``what does this [guide]line entail for the sentiment analysis product [they] were developing.``} During W1, UX5 shared with DS2 and DS4 that they developed \emph{``modules that characterize how different guidelines could be represented back in concrete examples... to help colleagues understand what these [AI fairness] guidelines would mean in an actual solution.’’} For example, based on their user experience research, UX5 documented the stakeholders who directly interacted with their system --- and those who might not directly interact with the systems but who might still be affected --- in a shareable document that was available across multiple AI teams in their organization. In the workshop, DS2 and DS4 both agreed with UX5 that these modules they created were extremely valuable to facilitate communications among team members in ways that attend to the real-world context for fairness issues.

\subsection{Piggybacking as a Tactic for Collaboration under Organizational Constraints} \label{piggybacking}

In parallel to ``bridging,'' we find that participants employed ``piggybacking’’ as a tactic to push fairness work forward in organizations that might not otherwise provide resources or incentives for fairness work. The ``piggybacking'' observed in our study took multiple forms: a) some participants piggybacked on related institutional processes, such as privacy impact assessments, to get buy-in for fairness work, b) some also positioned their work within the ``quantification'' culture of AI development to better communicate with technical roles on their teams. However, when sharing how these piggybacking tactics may have enabled them to conduct fairness work on cross-functional teams, participants also expressed their concerns around the limitations and compromised nature of these tactics.\looseness=-1

\subsubsection{Piggybacking on institutionalized procedures}
\label{privacy}

To address challenges in AI fairness work, some participants (DS3, PM1, UX4, SME1) shared their strategies around piggybacking on organization-wide mandatory privacy-related procedures (e.g., questionnaires, checklists) in order to raise awareness about AI fairness and put AI fairness efforts into practice. For example, DS3 developed a set of checklists to help the AI product teams reflect and assess if their AI features contained potential racial biases that might harm their users, but ran into challenges to incorporate these artifacts into the current day to day AI work. However, since {``\textit{team members and leadership are extremely cognizant about privacy}’’} at DS3’s company, the privacy team in the company had already developed a questionnaire called a \emph{``privacy impact assessment,’’} containing 10 privacy-related questions for all product teams to complete before launching any AI features or products. DS3 shared stories about how they built allyship with the privacy team and later piggybacked on the \emph{``privacy impact assessment’’} to promote their AI fairness effort:
%\begin{quote}
    \emph{``After multiple times getting rejected by our company to implement our fairness checklists, my teammate had this brilliant idea which was: What if we piggyback onto an existing process that already exists?… all we did was add an extra set of questions specifically concerning machine learning fairness to the privacy impact assessment… the beauty of that is that we don’t have to persuade people to fill out this form since they already have to fill out the larger privacy impact assessment in order to launch their products or features.''} (DS3)
%\end{quote}

Furthermore, DS3 told us that adding fairness-related questions to the privacy impact assessment helped to bring the awareness of AI fairness to DS3’s entire organization, as there were increasing amount of \emph{``people from other teams reaching out and saying that they want to work on fairness too.''} By doing so, participants were also able to build a coalition of team members and develop a network of allies within the organization who were committed to advancing fairness in the company's AI systems. 

Similarly, during the interview, PM1 brought up the concept of \emph{``change management''}---approaches to prepare for and support organizational changes. PM1 told us that \emph{``change management is very, very difficult for all of software practice, but especially with responsible AI, when we need to go the extra mile and explain why this is so important.''} Therefore, PM1 would \emph{``always start by adding little things to [their] privacy practice instead of creating an entirely new fairness thing... we don't have to evangelize about how important privacy is.''}

However, both DS3 and PM1 shared their desire to implement stand-alone fairness assessment procedures \emph{``eventually when there is more buy-in for AI fairness''} (PM1). DS3 shared that their current efforts around piggybacking on the privacy team might not be sustainable, and they wanted higher-level leadership to allocate more resources for AI fairness work. Furthermore, DS3 brought up the need for further exploring the trade-offs and complementarity between privacy and fairness work instead of smuggling fairness in with privacy assessments.

\subsubsection{Piggybacking on the quantification culture of AI development}
\label{quantification}

As mentioned in Section \ref{evaluation}, various disciplines may value different evaluation goals and methods for AI fairness. Current AI development culture still largely favors quantitative over qualitative methodologies and forms of evidence \cite{birhane2022values, ayling2022putting, Gebru2022Hierarchy}. As a result, many participants reported drawing on quantitative approaches in order to overcome communication barriers with team members in technical roles. 
For example, PM1 shared strategies around using a quantitative score to fit their AI fairness work into existing AI work metric systems and the \emph{``numerical culture of AI modeling work’’}: 
%\begin{quote}
    \emph{``It is always hard to convince the teams across organizations to accept something new, unless it is something that they are already familiar with [...] so we started using a scale of 1 to 10. If your [model’s] feature is not even showing any explainability or analysis around fairness then the score will be low. Engineers, they just don’t like low scores''} (PM1).
%\end{quote}
In other words, PM1 used these scores to make the value of working on AI fairness directly relevant to the data scientists on the team. During the workshops, another product manager, PM3, told us that when they are communicating with data scientists, \emph{``using [a fairness toolkit] to calculate numbers like demographic disparity made it much more straightforward for communicating with other data scientists about the impact of working on fairness.’’} Within the same workshop, UX6 concurred with PM3 by sharing that \emph{``having quantifiable fairness related scores,’’} was a commonly used strategy that worked well when collaborating with data scientists, and shared with us that they often change their communication strategy to include significant amount of quantitative data when working with data scientists on AI fairness.

However, many participants were aware of the potential pitfalls of using ``scores’’ and ``percentages’’ in AI fairness work. For example, PM2 told us that even though they would always \emph{``make sure to show something quantitative’’} when communicating with engineers and data scientists as a \emph{``trick to communicate,''} they don’t believe quantitative data is \emph{``necessarily the best way to represent the concept of fairness.''} During the workshop (W3), while PM3 shared how relying on quantification dramatically helped them with communicating with the AI development team, they also acknowledged that the numbers produced using fairness toolkits served as an (often inaccurate) proxy of the actual fairness harms that might be caused by their AI products, \emph{``essentially losing a lot of nuances of fairness’’} (PM3). As a result, while creating their ``ideal journey map,'' PM3 and others in their workshop shared the desire for better processes to help navigate the current AI development culture around prioritizing quantification to appropriately address socio-technical challenges that may require drawing on and integrating both quantitative and qualitative evidence of (un)fairness.

\subsection{Invisible Labor and Burdens in Cross-functional Collaboration} \label{invisible_labor}

While working towards more effective cross-functional collaboration, participants shared that their efforts during the collaborations were often invisible to their team members or leadership. Worse, sometimes other team members would hold unrealistic expectations of AI fairness work, creating additional burden and frustration for participants.\looseness=-1

Participants reported that one reason their ``bridging'' and ``piggybacking'' efforts encountered difficulties was that other team members did not always understand what AI fairness work actually involves. This contributed to under-recognition and unrealistic expectations. For example, during the workshop (W2), DS1, MLE1 and DS5, all from different organizations, had discussed how creating educational documentation or accessible visualizations to bridge the knowledge gaps among participants required substantial effort. MLE1 mentioned that updating the educational documentation to keep up with AI fairness research was also extremely time consuming, sometimes requiring several rounds of major iteration even within a week. The value of such extra efforts, however, was usually overlooked by their team leadership. As DS5 concurred with DS1 and MLE1 during the workshop session, 
%\begin{quote}
    \emph{``teams don’t understand the amount of work that goes into doing AI fairness work. It’s not quite as simple as pulling from a table and doing a statistical analysis. You have to put thoughts into making intentional analysis choices and think broadly about the socio-technical nature of AI fairness.} (DS5)\looseness=-1
%\end{quote}

Furthermore, participants shared that other team members, often technical roles who just started working on AI fairness, often held unrealistic expectations that collaboration around fairness would be a one-off effort, rather than an iterative, thoughtful process requiring long-term engagement. For example, UX5 shared that some colleagues expected design workshops around AI fairness (see \ref{understanding}) to be a \emph{``one time thing''} rather than a recurring activity---for instance, a data scientist on UX5's team asked \emph{``did[n't] we already do that last time and get the answer?”} when UX5 scheduled another design workshop around AI fairness. 

In addition, since AI fairness efforts were not reflected in most organizations' annual employee evaluations or other organizational incentives, individuals who were motivated to do fairness work often felt responsible for \emph{``holding the entire team accountable for the fairness work''} (MLE2), rather than having that accountability located in, for instance, ``ethics owners'' \cite{metcalf2019owning} or other organizational leadership. Similarly, MLE1 shared that they felt \emph{``frustrated by the attitude of [others] dodging responsibility for fairness work.”} However, MLE1 reported that they often felt uncomfortable, as a woman of color, \emph{``giv[ing] everyone a lecture on why fairness should be prioritized."} As a result, ``bridging'' work often falls onto a small number of team members who are self-motivated to do the work, yet who are not incentivized or supported in sustaining their efforts long term \cite[cf.][]{madaio2020co}. As RS2 shared, \emph{``engineers and data scientists just kind of leave the conversation behind, walking out of the meetings [...], and I have to spend more time to do most of the work to make sure the team can crystallize certain principles or certain best practices.''} 

As a result, participants across roles all desired better strategies to help team members better understand the iterative nature of AI fairness work, as well as to better advocate for and incentivize long-term collaboration across roles. When creating the ``ideal journey map'' during the workshop (W5), SWE2 and SME2 wanted team members to continuously address AI fairness across the lifecycle of AI development, instead of \emph{``discussing fairness almost like an empty gesture at the beginning of each season''} (SME2). In the same workshop, SME2 expressed their desire for organizations to cultivate a shared understanding among different roles regarding the continuous efforts required to address fairness in AI, as well as each role's unique contributions to collaboratively building more fair and responsible AI systems. Furthermore, participants repeatedly brought up the importance of changing the \emph{``education and training team members received''} (PM3) to set up a common grounding and expectations among members of their organization about the iterative, socio-technical nature of AI fairness practice. %, rather than leaving that to individuals to create and deliver. 

\section{Limitations}
% Our findings shed light on current collaboration practices, challenges, and needs among a set of practitioners who may be at the forefront of an emerging industry practice. 
As discussed in Section \ref{Methods}, similar to prior FAccT work studying AI fairness practices in industry \cite[e.g.,][]{holstein2019improving, madaio2020co, rakova2021responsible, deng2022exploring}, we recruited participants using a purposive sampling approach \cite{campbell2020purposive}. Although interview-based qualitative research does not necessarily make claims to generalizability \cite{small2022qualitative}, the participants we were able to recruit may have been limited by our positionality and the reach of our professional network. In addition, all of our participants self-identified as having already worked on addressing AI fairness issues in their AI products and services (and may thus lead to some self-selection bias), most of our participants worked at large technology companies, and the majority of our participants were located in the US or Europe. Future research should explore perspectives from people contributing to AI systems from outside technology companies, including non-profits, government agencies, and members of the public more generally; as well as perspectives from stakeholders outside of the U.S. and Europe. In addition, future research should explore perspectives from other roles involved in procuring and deploying AI applications, including roles focused on business outcomes (e.g., customer success, sales, marketing) to explore how work on fairness in AI is impacted by (or might resist) industry business logics or market incentives; and how community perspectives might be brought in conversation with members of AI teams from business roles. \looseness=-1

% integrating RAI work with cost.... 
\section{Discussion} \label{Discussion}

Cross-functional collaboration is critical in building more fair and responsible AI systems, yet is often absent or ineffective in industry AI fairness work. Through a series of interviews and workshops with industry practitioners, we have identified existing practices that practitioners have developed to support cross-functional collaboration. Based on our findings, we discuss opportunities for FAccT researchers and organizations to better support cross-functional collaboration in AI fairness work.

\subsection{Supporting Bridging Work in Cross-Functional Collaboration} \label{support_bridging}

Our findings surfaced critical bridging activities that practitioners use to overcome barriers to cross-functional collaboration (Section \ref{bridging}). How might researchers better support practitioners in bridging gaps in evaluation, understanding, and contextualization around AI fairness for their teams?

In our study, we found that practitioners spent significant effort navigating mismatches in goals and approaches for AI fairness evaluations across different disciplines (Section \ref{evaluation}). However, AI fairness methods and toolkits \cite[e.g.,][]{cabrera2019fairvis, bird2020fairlearn, FairlearnAPI, AIF360API} often focus primarily on supporting more technical roles in conducting quantitative analyses \cite{deng2022exploring, wong2022seeing, bauerle2022symphony}, despite rhetoric about enabling \textit{socio}-technical work \cite{wong2022seeing}. Hence, future research should explore mechanisms to support mixed-methods approaches to addressing fairness issues in AI. Drawing inspiration from recent HCI research that aims to empower collaboration between model- and user-facing roles in AI development \cite[cf.][]{subramonyam2021towards, lam2023model,wang2023designing}, future work could explore the development of new tools and processes to aid cross-functional teams in meaningfully integrating fairness-related insights from quantitative analyses (e.g., statistical disparities across subgroups) and qualitative data from user research (e.g., insights into perceptions of algorithmic harms among users or other groups impacted by a system). Such resources could support the sorts of cross-functional ``co-evaluation meetings'' discussed in Section \ref{evaluation}.\looseness=-1 

In addition, future research should explore ways to better support cross-functional teams in \textit{translating} evaluation results into actionable steps that appropriately account for the socio-technical complexities involved in a given use case or context. In our study, we found that participants from both technical roles (DS) and user-facing roles (UX) engaged in substantial work to translate and contextualize abstract AI fairness concepts in the documentation they created for those in other roles (Section \ref{contextualization}). We see opportunities to reduce the amount of translation work that falls on practitioners by developing more context- and usage-specific AI fairness guidelines and frameworks that highlight unique considerations for specific real-world applications, especially since resources and guidelines from industry and academia often serve as a starting point for practitioners to develop resources for their own teams and organizations \cite{amershi2019guidelines, Shneiderman2020HumanCenteredAI, varshney2019trustworthy, lewicki2023out}. For instance, to better support AI teams working on building chatbot services for health care, FAccT researchers and practitioners could design AI fairness guidelines covering how to assess and mitigate potential fairness issues (e.g., around sex, gender, and race) that might occur in health care applications \cite{cirillo2020sex, obermeyer2019dissecting}, such as dataset documentation tools specifically tailored for health data, such as Healthsheets \cite{rostamzadeh2022healthsheet}. More generally, this might involve developing tools or other resources to support this translation work on cross-functional teams.
%As recent guidance published by public sector agencies like \emph{``Blueprint for an AI Bill of Rights''} \cite{BillOfRight} and \emph{``AI Risk Management Framework''} \cite{AIRiskManageF} being recently released, future AI fairness guidelines should also update their relevant principles and strategies to reflect these regional governance and policies. 

%\ken{Commented out some text above for now, since it would need some more work for inclusion in the camera-ready (and it's not fully clear to me that including this is high priority). But feel free to restore and iterate upon it!}

%Our findings speak to the concrete strategies and procedures to 

Finally, we find that practitioners bridge their understanding and contextualize AI fairness concepts through discussion and deliberation across roles (Section \ref{understanding}). While prior work suggested the importance of creating spaces and boundary objects for data scientists and UX practitioners to try to integrate AI model- and and product-level considerations \cite{subramonyam2022solving, yildirim2022experienced, wang2023designing}, our findings highlighted that the socio-technical, contested nature of AI fairness in particular \cite{green2021contestation, barocas2016big} adds layers of complexity to the process of cross-functional communication and collaboration (Section \ref{understanding} and \ref{contextualization}). To this end, future research could draw inspiration from resources like Timelines \cite{wong2021timelines} and Value Cards \cite{shen2021value} to design processes for practitioners across roles to exchange perspectives and deliberate around AI fairness issues. In particular, based on our findings, these processes should scaffold practitioners in negotiating with each other about the problem formulation of fairness issues \cite[cf.][]{passi2019problem}, discussing the trade-offs among particular fairness metrics (or between quantitative and qualitative approaches to understanding fairness) \cite{friedler2021possibility, kleinberg2016inherent}, aligning disagreements around how to operationalize fairness in specific use cases \cite{obermeyer2019dissecting, raji2020saving}, and learning about how their AI products and services might impact end users' day to day experience when interacting with the AI systems \cite{shen2021everyday, devos2022toward, deng2023understanding}.

%\dhw{mark: new texts below, please review/revise}

%Finally, while we have observed active involvement from diverse roles in bridging activities, our findings suggest that not all roles are equally engaged in certain bridge activities. However, prior research has highlighted the importance of developing educational and training materials that extend beyond the technical aspects of AI fairness \cite{deng2022exploring, yildirim2023investigating}. Thus, our findings indicate the need for empowering non-technical roles to develop -- or co-develop with technical roles -- educational and onboarding materials for AI teams, to better address AI fairness issues that situated in specific contexts and applications.

\subsection{Thriving, Not Just Surviving: From Piggybacking to Re-shaping Organizational Culture}

In our study, we observed that, faced with a lack of resources and organizational buy-in around fairness work, practitioners often ``piggybacked'' on existing initiatives that have buy-in and support (Section \ref{piggybacking}). This tactic has previously been documented in other settings beyond AI fairness work, including environmental sustainability efforts and social justice work in industry \cite{Brown1998MakingBI, robinson2010save}. Researchers have argued that ``piggybacking'' effectively helped ideas and practices about environmental sustainability ``survive'' under conditions with limited organizational resources. However, in order for efforts around ethical work such as sustainability to ``thrive'' \cite{Miceli2021ThrivingNJ}, it is critical for practitioners to fundamentally ``reshape'' the organizational culture towards more sustainable values. % the ``long-term growth and resilience led by prioritizing sustainablity''. 

Our findings similarly point to the value and relevance of such ``piggybacking'' tactics for enacting collaboration in AI fairness work only \emph{as a starting point}. We believe there are still opportunities for practitioners working on AI fairness to go beyond ``piggybacking'' and enable broader ``reshaping'' \cite[cf.][]{wong2021tactics} of the organizational culture around AI fairness. In particular, drawing on work by Nafus and Sherman \cite{Nafus2014ThisOD}, Wong described ``tactics of soft resistance'' that UX practitioners employ to make the value of their work relevant to other roles, and to \emph{re-shape} their organization's culture, by rooting their resistance ``within a broader logic of the role of the market or the usefulness of technology'' \cite{wong2021tactics}. 

%\dhw{new texts below around piggybacking on institutionalized procedure. Could be potentially trimmed as the texts are quite long}

To this end, future research should explore opportunities to support practitioners in re-shaping their organizational culture, to help practitioners working on AI fairness ``thriv[e], not just surviv[e]'' \cite{Miceli2021ThrivingNJ} when navigating collaborations around AI fairness within the organizational constraints. For example, resonating with multiple participants' call to explore the complementarity between privacy and fairness (Section \ref{privacy}), future research could explore more in-depth synergies and interplay between privacy and fairness, especially when responsible AI guidelines list AI fairness and privacy side-by-side as part of the key components to enable ethical and responsible AI systems \cite{jobin2019global}. For example, Tahaei et al. studied how ``privacy champions''---individuals who strongly care about advocating for privacy---navigate challenges like internal prioritization tensions and limited tool support \cite{tahaei2021privacy}. Similarly, our research, aligned with prior work around industry AI fairness practices \cite{madaio2020co, rakova2021responsible, deng2023understanding, wang2023designing}, revealed comparable situations for ``fairness champions,'' those who individually promote fairness and often provide invisible labor to carry out AI fairness collaborations (As in Section \ref{invisible_labor}). Future FAccT research could explore methods for fostering communication and collaboration between roles like ``privacy champions'' and ``fairness champions,'' exploring possible synergies between AI fairness activities and privacy initiatives, along with other institutionalized processes.

However, it is important to acknowledge that privacy and security are still often de-prioritized by individual developers and organizations \cite{gutfleisch2022does, li2018coconut,bamberger2011privacy}. In particular, prior research indicated that the deprioritization of privacy is especially pronounced in smaller organizations or teams \cite{gutfleisch2022does}. Indeed, participants in our study who mentioned piggybacking on privacy procedure were all from large-size technology companies (with 25,000+ employees) that may already have more established privacy processes than smaller organizations in place. In addition, in their recent work ``Industry Unbound,'' Waldman demonstrated that current technology companies still have a long way to go around implementing privacy-aware values in substantive ways (i.e., as more than requirements to be adhered to) \cite{waldman2021industry}. To this end, future FAccT research and practice are needed to explore the following open questions: how might practitioners in smaller companies carry out AI fairness collaboration efforts when there are limited institutionalized procedures (such as privacy) to piggyback upon? How should industry practitioners working on building fair and responsible AI navigate the tension between achieving immediate, short-term results via piggybacking (which might result in performative actions \cite{tahaei2021privacy}) and enabling long-term, profound organizational change around value and culture?

Finally, practitioners in user- and product-facing roles in our study resorted to using metrics they felt were over-simplifications to communicate nuanced fairness-related aspects of their systems with quantitatively-oriented practitioners in model-facing roles. Future work is needed to enable user- and product-facing practitioners to effectively communicate socio-technical concepts such as fairness in ways that are legible and actionable by model developers, without sacrificing the socio-technical nuances in AI fairness. In particular, tying closely to the point around mixed-methods evaluation approach discussed in the previous section (See section \ref{support_bridging}), instead of simply relying on ``scores'' and ``percentages'' to make fairness issues relevant to model-facing roles, future FAccT research and practice could design better structures and processes to help incorporate qualitative notions of fairness into the current AI development process. 

For example, echoing insights from prior work studying the use of responsible AI toolkits and guidelines \cite{deng2022exploring, yildirim2023investigating}, future FAccT research and practice could explore extending the current model-building platforms (e.g., \cite{Jupyter, CoLab}) with contextual messages that bring in nuanced understandings around AI fairness from direct stakeholders, to integrate mixed-method, socio-technical AI fairness analysis into technical roles' existing AI working pipelines. We acknowledge that organizational change is a complex and time-intensive process, necessitating sustained efforts rather than being a sudden occurrence \cite{scott2015organizations}. However, it is our hope that, over time, tools and processes supporting mixed-method evaluation and development for AI fairness could potentially \textit{re-shape} the quantification culture around AI development that runs the risk of compromising the socio-technical, contested nature of AI fairness \cite{Gebru2022Hierarchy, selbst2019fairness, yurrita2023disentangling,forsythe2001studying}.

\subsection{Making Invisible Labor Visible and Valuable}

Throughout our findings, we see that carrying out the efforts of bridging and piggybacking often requires practitioners to go beyond their traditional job descriptions, devote additional time and effort, and take on emotional burdens (Section \ref{invisible_labor})---what Star and Strauss described as ``invisible work'' \cite{star1999layers}. This finding also draws a parallel with the recent journalism about the burnout problem in industry responsible AI work---due to the lack of appropriate recognition of their invisible work from colleagues and organizations, practitioners reported ``feeling undervalued, which can affect their mental health and lead to burnout'' \cite{RAI_burnout}. More recently, Wang et al. also surfaced the ``hidden work'' and ``emotional labor'' of UX practitioners when raising responsible AI issues in early-stage prototyping of AI applications, and how this work is often not recognized or valued by UX practitioners' managers or organizations, or leads others to view UX practitioners as a ``blocker'' of the design and development process \cite{wang2023designing}. We extend this prior work by highlighting how practitioners \emph{across roles} in AI teams (including UX practitioners) brought in their relevant expertise and perspectives to go beyond their job descriptions to enact the collaboration in AI fairness. 

Our findings identified the burden of practitioners' invisible labor when working towards more effective collaboration around AI fairness. Future research should thus explore processes and tools that help team members and organizations better recognize and value the efforts that enable fairness work. For instance, abstraction has been highlighted as an important skill for collaborating and communicating in software engineering and data analysis in cross-functional teams \cite{amershi2019software, liskov1987keynote, nahar2021collaboration}---although with the risk of losing the nuance of particular contexts \cite[cf.][]{selbst2019fairness,Hutchinson2022EvaluationGI}. However, in the context of collaboration on AI fairness work, these abstractions that were intended to facilitate conversations across roles often resulted in other team members not fully understanding and appreciating the labor hidden behind the efforts individuals invested in enabling the collaboration in AI fairness \ref{invisible_labor}. This finding is well-aligned with Kross and Guo's observations around how external clients failed to recognize the amount of data work done by data scientists while collaborating \cite{kross2021orienting}. In line with Wang et al.'s suggestion for supporting UX practitioners' hidden work \cite{wang2023designing}, our findings highlight the importance for organizations to recognize and incentivize the evolving roles of practitioners as ``translators,'' ``educators,'' and ``activist/advocates'' \cite[cf.][]{chivukula2021identity}. These emerging roles and responsibilities AI practitioners voluntarily take on, to some extent, blur the existing professional boundaries in the pursuit of more responsible and fair AI systems \cite{abbott2014system}. This raises the need for organizations to establish new educational programs, training, and even specific job descriptions for AI practitioners committed to the development of more equitable and responsible AI.

%In addition, an important step after recognizing the ``invisible work'' would be rewarding them. However, as repeatedly discussed in prior work, the efforts and the impact of ``invisible work'' are often hard to make clear or quantify \cite{star1999layers, wong2021tactics, wang2023designing} We suggest that FAccT researchers could draw inspiration from prior work attempting to quantify the invisible labour in crowdwork \cite[cf.][]{Toxtli2021QuantifyingTI} as well as the tools to track and report ``work tasks related to raising discussion and address[ing] values at work places'' that are often invisible to other team members and organizations \cite[cf.][]{wong2021using}. It is our hope that these new structures and tools could to some extent ameliorate practitioners' burnout induced by taking on under-recognized work.

Beyond recognizing emerging roles and responsibilities in AI fairness work, our finding suggested that companies need to proactively \emph{reward} this critical invisible labor that enables cross-functional collaboration in AI fairness. One concrete way to implement this is to include the time and efforts devoted to this invisible labor as part of companies' performance indicators (e.g., Key Performance Indicators (KPIs), or Objectives and Key Results (OKRs)) in order to incentivize collaborative fairness efforts---although adapting KPIs for fairness or responsible as a whole brings with it a host of tensions and contradictions around quantifying unobservable phenomena such as fairness \cite{madaio2020need, lee2021formalising}. In doing so, organizations could enhance the visibility and the mutual understanding among roles about the efforts to carry out collaboration on AI fairness. Extending the implications from Wong et al. around better supporting UX designers' values work by assigning them roles like ``responsible AI expert,'' \cite{wong2021tactics} organizations could also explore formalizing more teams and roles for fairness work to empower practitioners across roles who currently go beyond their traditional job descriptions to facilitate cross-functional collaboration for building more fair AI systems.

%%To support the mutual understanding among roles about the efforts to carry out collaboration on AI fairness, future work could design interactive interfaces or visualizations offering options for practitioners across roles to recognize the often laborious work that goes into the complex considerations and tradeoffs involved in fairness work, such as deciding on appropriate fairness metrics or larger analysis decisions when conducting disaggregated evaluations of fairness \cite{Barocas2021DesigningDE, madaio2022assessing}, or the arduous processes that enable the clear presentation of an accessible visualization from data scientists (Section \ref{invisible_labor}). 

\section{Conclusion}
In this research, we sought to better understand current strategies and challenges for cross-functional collaboration around fairness in AI, in order to identify opportunities to support more effective collaboration. Through a series of interviews and workshops with industry practitioners across a range of roles and companies, we found that practitioners engaged in critical, yet under-recognized \emph{bridging} work to help teams overcome key barriers to cross-functional collaboration around AI fairness. In addition, given organizational constraints, practitioners often \emph{piggybacked} on existing initiatives and corporate rhetorics to enable fairness efforts. Overall, we hope that this work can
(1) increase awareness among the FAccT community around the strategies and tactics that industry practitioners currently employ to facilitate collaborative AI fairness work; and (2) offer directions for future FAccT research and practice to better support cross-functional collaboration.

%\ken{Rather than spending too much time on another recap (which is already covered by the Discussion), perhaps close with a brief sentence or two about what you hope this work will help the FAccT community do.}

%%
%% The acknowledgments section is defined using the "acks" environment
%% (and NOT an unnumbered section). This ensures the proper
%% identification of the section in the article metadata, and the
%% consistent spelling of the heading.
\begin{acks}
This work was supported by the National Science Foundation (NSF) program on Fairness in AI in collaboration with Amazon under Award No. IIS-2040942, an award from Cisco Research, and an award from the Jacobs Foundation. We would like to thank Miro Dudik, Jenn Wortman Vaughan
, and Hanna Wallach for their feedback on the early stages of this project. Special thanks to our anonymous reviewers and to all participating industry practitioners for making this work possible.
\end{acks}

\clearpage

%%
%% The next two lines define the bibliography style to be used, and
%% the bibliography file.
\bibliographystyle{ACM-Reference-Format}
\bibliography{citation}

%%% -*-BibTeX-*-
%%% Do NOT edit. File created by BibTeX with style
%%% ACM-Reference-Format-Journals [18-Jan-2012].

\begin{thebibliography}{99}

%%% ====================================================================
%%% NOTE TO THE USER: you can override these defaults by providing
%%% customized versions of any of these macros before the \bibliography
%%% command.  Each of them MUST provide its own final punctuation,
%%% except for \shownote{}, \showDOI{}, and \showURL{}.  The latter two
%%% do not use final punctuation, in order to avoid confusing it with
%%% the Web address.
%%%
%%% To suppress output of a particular field, define its macro to expand
%%% to an empty string, or better, \unskip, like this:
%%%
%%% \newcommand{\showDOI}[1]{\unskip}   % LaTeX syntax
%%%
%%% \def \showDOI #1{\unskip}           % plain TeX syntax
%%%
%%% ====================================================================

\ifx \showCODEN    \undefined \def \showCODEN     #1{\unskip}     \fi
\ifx \showDOI      \undefined \def \showDOI       #1{#1}\fi
\ifx \showISBNx    \undefined \def \showISBNx     #1{\unskip}     \fi
\ifx \showISBNxiii \undefined \def \showISBNxiii  #1{\unskip}     \fi
\ifx \showISSN     \undefined \def \showISSN      #1{\unskip}     \fi
\ifx \showLCCN     \undefined \def \showLCCN      #1{\unskip}     \fi
\ifx \shownote     \undefined \def \shownote      #1{#1}          \fi
\ifx \showarticletitle \undefined \def \showarticletitle #1{#1}   \fi
\ifx \showURL      \undefined \def \showURL       {\relax}        \fi
% The following commands are used for tagged output and should be
% invisible to TeX
\providecommand\bibfield[2]{#2}
\providecommand\bibinfo[2]{#2}
\providecommand\natexlab[1]{#1}
\providecommand\showeprint[2][]{arXiv:#2}

\bibitem[Abbott(2014)]%
        {abbott2014system}
\bibfield{author}{\bibinfo{person}{Andrew Abbott}.}
  \bibinfo{year}{2014}\natexlab{}.
\newblock \bibinfo{booktitle}{\emph{The system of professions: An essay on the
  division of expert labor}}.
\newblock \bibinfo{publisher}{University of Chicago press}.
\newblock


\bibitem[AI(2021)]%
        {AIF360API}
\bibfield{author}{\bibinfo{person}{IBM Resaerch~Trusted AI}.}
  \bibinfo{year}{2021}\natexlab{}.
\newblock \showarticletitle{AIF360 API}.
\newblock  (\bibinfo{year}{2021}).
\newblock
\urldef\tempurl%
\url{https://aif360.mybluemix.net/}
\showURL{%
\tempurl}


\bibitem[Almahmoud et~al\mbox{.}(2021)]%
        {almahmoud2021teams}
\bibfield{author}{\bibinfo{person}{Jumana Almahmoud}, \bibinfo{person}{Robert
  DeLine}, {and} \bibinfo{person}{Steven~M Drucker}.}
  \bibinfo{year}{2021}\natexlab{}.
\newblock \showarticletitle{How Teams Communicate about the Quality of ML
  Models: A Case Study at an International Technology Company}.
\newblock \bibinfo{journal}{\emph{Proceedings of the ACM on Human-Computer
  Interaction}} \bibinfo{volume}{5}, \bibinfo{number}{GROUP}
  (\bibinfo{year}{2021}), \bibinfo{pages}{1--24}.
\newblock


\bibitem[Amershi et~al\mbox{.}(2019a)]%
        {amershi2019software}
\bibfield{author}{\bibinfo{person}{Saleema Amershi}, \bibinfo{person}{Andrew
  Begel}, \bibinfo{person}{Christian Bird}, \bibinfo{person}{Robert DeLine},
  \bibinfo{person}{Harald Gall}, \bibinfo{person}{Ece Kamar},
  \bibinfo{person}{Nachiappan Nagappan}, \bibinfo{person}{Besmira Nushi}, {and}
  \bibinfo{person}{Thomas Zimmermann}.} \bibinfo{year}{2019}\natexlab{a}.
\newblock \showarticletitle{Software engineering for machine learning: A case
  study}. In \bibinfo{booktitle}{\emph{2019 IEEE/ACM 41st International
  Conference on Software Engineering: Software Engineering in Practice
  (ICSE-SEIP)}}. IEEE, \bibinfo{pages}{291--300}.
\newblock


\bibitem[Amershi et~al\mbox{.}(2019b)]%
        {amershi2019guidelines}
\bibfield{author}{\bibinfo{person}{Saleema Amershi}, \bibinfo{person}{Dan
  Weld}, \bibinfo{person}{Mihaela Vorvoreanu}, \bibinfo{person}{Adam Fourney},
  \bibinfo{person}{Besmira Nushi}, \bibinfo{person}{Penny Collisson},
  \bibinfo{person}{Jina Suh}, \bibinfo{person}{Shamsi Iqbal},
  \bibinfo{person}{Paul~N Bennett}, \bibinfo{person}{Kori Inkpen},
  {et~al\mbox{.}}} \bibinfo{year}{2019}\natexlab{b}.
\newblock \showarticletitle{Guidelines for human-AI interaction}. In
  \bibinfo{booktitle}{\emph{Proceedings of the 2019 chi conference on human
  factors in computing systems}}. \bibinfo{pages}{1--13}.
\newblock


\bibitem[Ayling and Chapman(2022)]%
        {ayling2022putting}
\bibfield{author}{\bibinfo{person}{Jacqui Ayling} {and}
  \bibinfo{person}{Adriane Chapman}.} \bibinfo{year}{2022}\natexlab{}.
\newblock \showarticletitle{Putting AI ethics to work: are the tools fit for
  purpose?}
\newblock \bibinfo{journal}{\emph{AI and Ethics}} \bibinfo{volume}{2},
  \bibinfo{number}{3} (\bibinfo{year}{2022}), \bibinfo{pages}{405--429}.
\newblock


\bibitem[Bamberger and Mulligan(2011)]%
        {bamberger2011privacy}
\bibfield{author}{\bibinfo{person}{Kenneth~A Bamberger} {and}
  \bibinfo{person}{Deirdre~K Mulligan}.} \bibinfo{year}{2011}\natexlab{}.
\newblock \showarticletitle{Privacy on the Books and on the Ground}.
\newblock \bibinfo{journal}{\emph{Stanford Law Review}} (\bibinfo{year}{2011}),
  \bibinfo{pages}{247--315}.
\newblock


\bibitem[Barocas and Selbst(2016)]%
        {barocas2016big}
\bibfield{author}{\bibinfo{person}{Solon Barocas} {and}
  \bibinfo{person}{Andrew~D Selbst}.} \bibinfo{year}{2016}\natexlab{}.
\newblock \showarticletitle{Big data's disparate impact}.
\newblock \bibinfo{journal}{\emph{Calif. L. Rev.}}  \bibinfo{volume}{104}
  (\bibinfo{year}{2016}), \bibinfo{pages}{671}.
\newblock


\bibitem[B{\"a}uerle et~al\mbox{.}(2022)]%
        {bauerle2022symphony}
\bibfield{author}{\bibinfo{person}{Alex B{\"a}uerle},
  \bibinfo{person}{{\'A}ngel~Alexander Cabrera}, \bibinfo{person}{Fred Hohman},
  \bibinfo{person}{Megan Maher}, \bibinfo{person}{David Koski},
  \bibinfo{person}{Xavier Suau}, \bibinfo{person}{Titus Barik}, {and}
  \bibinfo{person}{Dominik Moritz}.} \bibinfo{year}{2022}\natexlab{}.
\newblock \showarticletitle{Symphony: Composing interactive interfaces for
  machine learning}. In \bibinfo{booktitle}{\emph{Proceedings of the 2022 CHI
  Conference on Human Factors in Computing Systems}}. \bibinfo{pages}{1--14}.
\newblock


\bibitem[Bird(2020)]%
        {FairlearnAPI}
\bibfield{author}{\bibinfo{person}{Sarah Bird}.}
  \bibinfo{year}{2020}\natexlab{}.
\newblock \bibinfo{title}{Fairlearn API}.
\newblock
\newblock
\urldef\tempurl%
\url{https://fairlearn.github.io/v0.5.0/api_reference/fairlearn.datasets.html}
\showURL{%
\tempurl}


\bibitem[Bird et~al\mbox{.}(2020)]%
        {bird2020fairlearn}
\bibfield{author}{\bibinfo{person}{Sarah Bird}, \bibinfo{person}{Miro Dudík},
  \bibinfo{person}{Richard Edgar}, \bibinfo{person}{Brandon Horn},
  \bibinfo{person}{Roman Lutz}, \bibinfo{person}{Vanessa Milan},
  \bibinfo{person}{Mehrnoosh Sameki}, \bibinfo{person}{Hanna Wallach}, {and}
  \bibinfo{person}{Kathleen Walker}.} \bibinfo{year}{2020}\natexlab{}.
\newblock \bibinfo{booktitle}{\emph{Fairlearn: A toolkit for assessing and
  improving fairness in AI}}.
\newblock \bibinfo{type}{{T}echnical {R}eport} MSR-TR-2020-32.
  \bibinfo{institution}{Microsoft}.
\newblock
\urldef\tempurl%
\url{https://www.microsoft.com/en-us/research/publication/fairlearn-a-toolkit-for-assessing-and-improving-fairness-in-ai/}
\showURL{%
\tempurl}


\bibitem[Birhane et~al\mbox{.}(2022)]%
        {birhane2022values}
\bibfield{author}{\bibinfo{person}{Abeba Birhane}, \bibinfo{person}{Pratyusha
  Kalluri}, \bibinfo{person}{Dallas Card}, \bibinfo{person}{William Agnew},
  \bibinfo{person}{Ravit Dotan}, {and} \bibinfo{person}{Michelle Bao}.}
  \bibinfo{year}{2022}\natexlab{}.
\newblock \showarticletitle{The values encoded in machine learning research}.
  In \bibinfo{booktitle}{\emph{2022 ACM Conference on Fairness, Accountability,
  and Transparency}}. \bibinfo{pages}{173--184}.
\newblock


\bibitem[Blodgett et~al\mbox{.}(2020)]%
        {blodgett2020language}
\bibfield{author}{\bibinfo{person}{Su~Lin Blodgett}, \bibinfo{person}{Solon
  Barocas}, \bibinfo{person}{Hal Daum{\'e}~III}, {and} \bibinfo{person}{Hanna
  Wallach}.} \bibinfo{year}{2020}\natexlab{}.
\newblock \showarticletitle{Language (technology) is power: A critical survey
  of" bias" in nlp}.
\newblock \bibinfo{journal}{\emph{arXiv preprint arXiv:2005.14050}}
  (\bibinfo{year}{2020}).
\newblock


\bibitem[Braun and Clarke(2006)]%
        {Braun2006-ts}
\bibfield{author}{\bibinfo{person}{Virginia Braun} {and}
  \bibinfo{person}{Victoria Clarke}.} \bibinfo{year}{2006}\natexlab{}.
\newblock \showarticletitle{Using thematic analysis in psychology}.
\newblock \bibinfo{journal}{\emph{Qual. Res. Psychol.}} \bibinfo{volume}{3},
  \bibinfo{number}{2} (\bibinfo{date}{Jan.} \bibinfo{year}{2006}),
  \bibinfo{pages}{77--101}.
\newblock


\bibitem[Brown and Larson(1998)]%
        {Brown1998MakingBI}
\bibfield{author}{\bibinfo{person}{Howard Brown} {and}
  \bibinfo{person}{Timothy~J. Larson}.} \bibinfo{year}{1998}\natexlab{}.
\newblock \showarticletitle{Making business integration work: A survival
  strategy for EHS managers}.
\newblock \bibinfo{journal}{\emph{Environmental Quality Management}}
  \bibinfo{volume}{7} (\bibinfo{year}{1998}), \bibinfo{pages}{1--8}.
\newblock


\bibitem[Burnie(2016)]%
        {burnie2016piggybacking}
\bibfield{author}{\bibinfo{person}{Helen Burnie}.}
  \bibinfo{year}{2016}\natexlab{}.
\newblock \showarticletitle{Piggybacking on sustainability}.
\newblock \bibinfo{journal}{\emph{Practical Literacy: The Early and Primary
  Years}} \bibinfo{volume}{21}, \bibinfo{number}{3} (\bibinfo{year}{2016}),
  \bibinfo{pages}{35--38}.
\newblock


\bibitem[Cabrera et~al\mbox{.}(2021)]%
        {cabrera2021discovering}
\bibfield{author}{\bibinfo{person}{{\'A}ngel~Alexander Cabrera},
  \bibinfo{person}{Abraham~J Druck}, \bibinfo{person}{Jason~I Hong}, {and}
  \bibinfo{person}{Adam Perer}.} \bibinfo{year}{2021}\natexlab{}.
\newblock \showarticletitle{Discovering and validating ai errors with
  crowdsourced failure reports}.
\newblock \bibinfo{journal}{\emph{Proceedings of the ACM on Human-Computer
  Interaction}} \bibinfo{volume}{5}, \bibinfo{number}{CSCW2}
  (\bibinfo{year}{2021}), \bibinfo{pages}{1--22}.
\newblock


\bibitem[Cabrera et~al\mbox{.}(2019)]%
        {cabrera2019fairvis}
\bibfield{author}{\bibinfo{person}{{\'A}ngel~Alexander Cabrera},
  \bibinfo{person}{Will Epperson}, \bibinfo{person}{Fred Hohman},
  \bibinfo{person}{Minsuk Kahng}, \bibinfo{person}{Jamie Morgenstern}, {and}
  \bibinfo{person}{Duen~Horng Chau}.} \bibinfo{year}{2019}\natexlab{}.
\newblock \showarticletitle{FairVis: Visual analytics for discovering
  intersectional bias in machine learning}. In \bibinfo{booktitle}{\emph{2019
  IEEE Conference on Visual Analytics Science and Technology (VAST)}}. IEEE,
  \bibinfo{pages}{46--56}.
\newblock


\bibitem[Cabrera et~al\mbox{.}(2022)]%
        {cabrera2022did}
\bibfield{author}{\bibinfo{person}{{\'A}ngel~Alexander Cabrera},
  \bibinfo{person}{Marco~Tulio Ribeiro}, \bibinfo{person}{Bongshin Lee},
  \bibinfo{person}{Rob DeLine}, \bibinfo{person}{Adam Perer}, {and}
  \bibinfo{person}{Steven~M Drucker}.} \bibinfo{year}{2022}\natexlab{}.
\newblock \showarticletitle{What Did My AI Learn? How Data Scientists Make
  Sense of Model Behavior}.
\newblock \bibinfo{journal}{\emph{ACM Transactions on Computer-Human
  Interaction}} (\bibinfo{year}{2022}).
\newblock


\bibitem[Campbell et~al\mbox{.}(2020)]%
        {campbell2020purposive}
\bibfield{author}{\bibinfo{person}{Steve Campbell}, \bibinfo{person}{Melanie
  Greenwood}, \bibinfo{person}{Sarah Prior}, \bibinfo{person}{Toniele Shearer},
  \bibinfo{person}{Kerrie Walkem}, \bibinfo{person}{Sarah Young},
  \bibinfo{person}{Danielle Bywaters}, {and} \bibinfo{person}{Kim Walker}.}
  \bibinfo{year}{2020}\natexlab{}.
\newblock \showarticletitle{Purposive sampling: complex or simple? Research
  case examples}.
\newblock \bibinfo{journal}{\emph{Journal of research in Nursing}}
  \bibinfo{volume}{25}, \bibinfo{number}{8} (\bibinfo{year}{2020}),
  \bibinfo{pages}{652--661}.
\newblock


\bibitem[Chivukula et~al\mbox{.}(2021)]%
        {chivukula2021identity}
\bibfield{author}{\bibinfo{person}{Shruthi~Sai Chivukula},
  \bibinfo{person}{Aiza Hasib}, \bibinfo{person}{Ziqing Li},
  \bibinfo{person}{Jingle Chen}, {and} \bibinfo{person}{Colin~M Gray}.}
  \bibinfo{year}{2021}\natexlab{}.
\newblock \showarticletitle{Identity Claims that Underlie Ethical Awareness and
  Action}. In \bibinfo{booktitle}{\emph{Proceedings of the 2021 CHI Conference
  on Human Factors in Computing Systems}}. \bibinfo{pages}{1--13}.
\newblock


\bibitem[Cirillo et~al\mbox{.}(2020)]%
        {cirillo2020sex}
\bibfield{author}{\bibinfo{person}{Davide Cirillo}, \bibinfo{person}{Silvina
  Catuara-Solarz}, \bibinfo{person}{Czuee Morey}, \bibinfo{person}{Emre Guney},
  \bibinfo{person}{Laia Subirats}, \bibinfo{person}{Simona Mellino},
  \bibinfo{person}{Annalisa Gigante}, \bibinfo{person}{Alfonso Valencia},
  \bibinfo{person}{Mar{\'\i}a~Jos{\'e} Rementeria},
  \bibinfo{person}{Antonella~Santuccione Chadha}, {et~al\mbox{.}}}
  \bibinfo{year}{2020}\natexlab{}.
\newblock \showarticletitle{Sex and gender differences and biases in artificial
  intelligence for biomedicine and healthcare}.
\newblock \bibinfo{journal}{\emph{NPJ digital medicine}} \bibinfo{volume}{3},
  \bibinfo{number}{1} (\bibinfo{year}{2020}), \bibinfo{pages}{81}.
\newblock


\bibitem[Colab(2020)]%
        {CoLab}
\bibfield{author}{\bibinfo{person}{Colab}.} \bibinfo{year}{2020}\natexlab{}.
\newblock \bibinfo{title}{Welcome To Colaboratory}.
\newblock
\newblock
\urldef\tempurl%
\url{https://colab.research.google.com/}
\showURL{%
\tempurl}


\bibitem[Cramer et~al\mbox{.}(2018)]%
        {cramer2018assessing}
\bibfield{author}{\bibinfo{person}{Henriette Cramer}, \bibinfo{person}{Jean
  Garcia-Gathright}, \bibinfo{person}{Aaron Springer}, {and}
  \bibinfo{person}{Sravana Reddy}.} \bibinfo{year}{2018}\natexlab{}.
\newblock \showarticletitle{Assessing and addressing algorithmic bias in
  practice}.
\newblock \bibinfo{journal}{\emph{Interactions}} \bibinfo{volume}{25},
  \bibinfo{number}{6} (\bibinfo{year}{2018}), \bibinfo{pages}{58--63}.
\newblock


\bibitem[Crawford(2017)]%
        {crawford2017trouble}
\bibfield{author}{\bibinfo{person}{Kate Crawford}.}
  \bibinfo{year}{2017}\natexlab{}.
\newblock \showarticletitle{The trouble with bias. keynote at neurips}.
\newblock  (\bibinfo{year}{2017}).
\newblock


\bibitem[Deng et~al\mbox{.}(2023)]%
        {deng2023understanding}
\bibfield{author}{\bibinfo{person}{Wesley~Hanwen Deng}, \bibinfo{person}{Boyuan
  Guo}, \bibinfo{person}{Alicia Devrio}, \bibinfo{person}{Hong Shen},
  \bibinfo{person}{Motahhare Eslami}, {and} \bibinfo{person}{Kenneth
  Holstein}.} \bibinfo{year}{2023}\natexlab{}.
\newblock \showarticletitle{Understanding Practices, Challenges, and
  Opportunities for User-Engaged Algorithm Auditing in Industry Practice}. In
  \bibinfo{booktitle}{\emph{Proceedings of the 2023 CHI Conference on Human
  Factors in Computing Systems}}. \bibinfo{pages}{1--18}.
\newblock


\bibitem[Deng et~al\mbox{.}(2022)]%
        {deng2022exploring}
\bibfield{author}{\bibinfo{person}{Wesley~Hanwen Deng}, \bibinfo{person}{Manish
  Nagireddy}, \bibinfo{person}{Michelle Seng~Ah Lee}, \bibinfo{person}{Jatinder
  Singh}, \bibinfo{person}{Zhiwei~Steven Wu}, \bibinfo{person}{Kenneth
  Holstein}, {and} \bibinfo{person}{Haiyi Zhu}.}
  \bibinfo{year}{2022}\natexlab{}.
\newblock \showarticletitle{Exploring how machine learning practitioners (try
  to) use fairness toolkits}. In \bibinfo{booktitle}{\emph{2022 ACM Conference
  on Fairness, Accountability, and Transparency}}. \bibinfo{pages}{473--484}.
\newblock


\bibitem[DeVos et~al\mbox{.}(2022)]%
        {devos2022toward}
\bibfield{author}{\bibinfo{person}{Alicia DeVos}, \bibinfo{person}{Aditi
  Dhabalia}, \bibinfo{person}{Hong Shen}, \bibinfo{person}{Kenneth Holstein},
  {and} \bibinfo{person}{Motahhare Eslami}.} \bibinfo{year}{2022}\natexlab{}.
\newblock \showarticletitle{Toward User-Driven Algorithm Auditing:
  Investigating users’ strategies for uncovering harmful algorithmic
  behavior}. In \bibinfo{booktitle}{\emph{Proceedings of the 2022 CHI
  Conference on Human Factors in Computing Systems}}. \bibinfo{pages}{1--19}.
\newblock


\bibitem[di~Miceli et~al\mbox{.}(2021)]%
        {Miceli2021ThrivingNJ}
\bibfield{author}{\bibinfo{person}{Andrea di Miceli}, \bibinfo{person}{Birgit
  Hagen}, \bibinfo{person}{Maria~Pia Riccardi}, \bibinfo{person}{Francesco
  Sotti}, {and} \bibinfo{person}{Davide Settembre-Blundo}.}
  \bibinfo{year}{2021}\natexlab{}.
\newblock \showarticletitle{Thriving, Not Just Surviving in Changing Times: How
  Sustainability, Agility and Digitalization Intertwine with Organizational
  Resilience}.
\newblock \bibinfo{journal}{\emph{Sustainability}} (\bibinfo{year}{2021}).
\newblock


\bibitem[Evenson(2006)]%
        {evenson2006directed}
\bibfield{author}{\bibinfo{person}{Shelley Evenson}.}
  \bibinfo{year}{2006}\natexlab{}.
\newblock \showarticletitle{Directed storytelling: Interpreting experience for
  design}.
\newblock \bibinfo{journal}{\emph{Design Studies: Theory and research in
  graphic design}} (\bibinfo{year}{2006}), \bibinfo{pages}{231--240}.
\newblock


\bibitem[EY(2021)]%
        {EYTrust}
\bibfield{author}{\bibinfo{person}{EY}.} \bibinfo{year}{2021}\natexlab{}.
\newblock \showarticletitle{EY's Trust Score: Defining business priorities for
  long-term value}.
\newblock  (\bibinfo{year}{2021}).
\newblock
\urldef\tempurl%
\url{https://www.ey.com/en_us/consulting/trusted-ai-platform}
\showURL{%
\tempurl}


\bibitem[Forsythe(2001)]%
        {forsythe2001studying}
\bibfield{author}{\bibinfo{person}{Diana Forsythe}.}
  \bibinfo{year}{2001}\natexlab{}.
\newblock \bibinfo{booktitle}{\emph{Studying those who study us: An
  anthropologist in the world of artificial intelligence}}.
\newblock \bibinfo{publisher}{Stanford University Press}.
\newblock


\bibitem[Friedler et~al\mbox{.}(2021)]%
        {friedler2021possibility}
\bibfield{author}{\bibinfo{person}{Sorelle~A Friedler}, \bibinfo{person}{Carlos
  Scheidegger}, {and} \bibinfo{person}{Suresh Venkatasubramanian}.}
  \bibinfo{year}{2021}\natexlab{}.
\newblock \showarticletitle{The (im) possibility of fairness: Different value
  systems require different mechanisms for fair decision making}.
\newblock \bibinfo{journal}{\emph{Commun. ACM}} \bibinfo{volume}{64},
  \bibinfo{number}{4} (\bibinfo{year}{2021}), \bibinfo{pages}{136--143}.
\newblock


\bibitem[Gebru(2021)]%
        {Gebru2022Hierarchy}
\bibfield{author}{\bibinfo{person}{Timnit Gebru}.}
  \bibinfo{year}{2021}\natexlab{}.
\newblock \bibinfo{title}{Hierarchy of Knowledge in Machine Learning and
  Related Fields and Its Consequences}.
\newblock
\newblock
\urldef\tempurl%
\url{https://www.youtube.com/watch?v=OL3DowBM9uc}
\showURL{%
\tempurl}


\bibitem[Gray et~al\mbox{.}(2010)]%
        {gray2010gamestorming}
\bibfield{author}{\bibinfo{person}{Dave Gray}, \bibinfo{person}{Sunni Brown},
  {and} \bibinfo{person}{James Macanufo}.} \bibinfo{year}{2010}\natexlab{}.
\newblock \bibinfo{booktitle}{\emph{Gamestorming: A playbook for innovators,
  rulebreakers, and changemakers}}.
\newblock \bibinfo{publisher}{" O'Reilly Media, Inc."}.
\newblock


\bibitem[Green(2021)]%
        {green2021contestation}
\bibfield{author}{\bibinfo{person}{Ben Green}.}
  \bibinfo{year}{2021}\natexlab{}.
\newblock \showarticletitle{The contestation of tech ethics: A sociotechnical
  approach to technology ethics in practice}.
\newblock \bibinfo{journal}{\emph{Journal of Social Computing}}
  \bibinfo{volume}{2}, \bibinfo{number}{3} (\bibinfo{year}{2021}),
  \bibinfo{pages}{209--225}.
\newblock


\bibitem[Gutfleisch et~al\mbox{.}(2022)]%
        {gutfleisch2022does}
\bibfield{author}{\bibinfo{person}{Marco Gutfleisch}, \bibinfo{person}{Jan~H
  Klemmer}, \bibinfo{person}{Niklas Busch}, \bibinfo{person}{Yasemin Acar},
  \bibinfo{person}{M~Angela Sasse}, {and} \bibinfo{person}{Sascha Fahl}.}
  \bibinfo{year}{2022}\natexlab{}.
\newblock \showarticletitle{How does usable security (not) end up in software
  products? results from a qualitative interview study}. In
  \bibinfo{booktitle}{\emph{43rd IEEE Symposium on Security and Privacy, IEEE
  S\&P}}. \bibinfo{pages}{22--26}.
\newblock


\bibitem[Heikkilä(2022)]%
        {RAI_burnout}
\bibfield{author}{\bibinfo{person}{Melissa Heikkilä}.}
  \bibinfo{year}{2022}\natexlab{}.
\newblock \bibinfo{title}{Responsible AI has a burnout problem}.
\newblock
\newblock
\urldef\tempurl%
\url{https://www.technologyreview.com/2022/10/28/1062332/responsible-ai-has-a-burnout-problem/}
\showURL{%
\tempurl}


\bibitem[Henke et~al\mbox{.}(1993)]%
        {Henke1993CrossFunctionalTG}
\bibfield{author}{\bibinfo{person}{John~W. Henke}, \bibinfo{person}{A.~Richard
  Krachenberg}, {and} \bibinfo{person}{Thomas~F. Lyons}.}
  \bibinfo{year}{1993}\natexlab{}.
\newblock \showarticletitle{Cross-Functional Teams: Good Concept, Poor
  Implementation!}
\newblock \bibinfo{journal}{\emph{Journal of Product Innovation Management}}
  \bibinfo{volume}{10} (\bibinfo{year}{1993}), \bibinfo{pages}{216--229}.
\newblock


\bibitem[Holstein et~al\mbox{.}(2019)]%
        {holstein2019improving}
\bibfield{author}{\bibinfo{person}{Kenneth Holstein}, \bibinfo{person}{Jennifer
  Wortman~Vaughan}, \bibinfo{person}{Hal Daum{\'e}~III}, \bibinfo{person}{Miro
  Dudik}, {and} \bibinfo{person}{Hanna Wallach}.}
  \bibinfo{year}{2019}\natexlab{}.
\newblock \showarticletitle{Improving fairness in machine learning systems:
  What do industry practitioners need?}. In
  \bibinfo{booktitle}{\emph{Proceedings of the 2019 CHI Conference on Human
  Factors in Computing Systems}}. \bibinfo{pages}{1--16}.
\newblock


\bibitem[Hutchinson et~al\mbox{.}(2022)]%
        {Hutchinson2022EvaluationGI}
\bibfield{author}{\bibinfo{person}{Benton~C. Hutchinson},
  \bibinfo{person}{Negar Rostamzadeh}, \bibinfo{person}{Christina Greer},
  \bibinfo{person}{Katherine Heller}, {and} \bibinfo{person}{Vinodkumar
  Prabhakaran}.} \bibinfo{year}{2022}\natexlab{}.
\newblock \showarticletitle{Evaluation Gaps in Machine Learning Practice}.
\newblock \bibinfo{journal}{\emph{2022 ACM Conference on Fairness,
  Accountability, and Transparency}} (\bibinfo{year}{2022}).
\newblock


\bibitem[Jobin et~al\mbox{.}(2019)]%
        {jobin2019global}
\bibfield{author}{\bibinfo{person}{Anna Jobin}, \bibinfo{person}{Marcello
  Ienca}, {and} \bibinfo{person}{Effy Vayena}.}
  \bibinfo{year}{2019}\natexlab{}.
\newblock \showarticletitle{The global landscape of AI ethics guidelines}.
\newblock \bibinfo{journal}{\emph{Nature Machine Intelligence}}
  \bibinfo{volume}{1}, \bibinfo{number}{9} (\bibinfo{year}{2019}),
  \bibinfo{pages}{389--399}.
\newblock


\bibitem[Jupyter(2020)]%
        {Jupyter}
\bibfield{author}{\bibinfo{person}{Jupyter}.} \bibinfo{year}{2020}\natexlab{}.
\newblock \bibinfo{title}{Jypyter: Free software, open standards, and web
  services for interactive computing across all programming languages}.
\newblock
\newblock
\urldef\tempurl%
\url{https://jupyter.org/}
\showURL{%
\tempurl}


\bibitem[Kahn(1996)]%
        {kahn1996interdepartmental}
\bibfield{author}{\bibinfo{person}{Kenneth~B Kahn}.}
  \bibinfo{year}{1996}\natexlab{}.
\newblock \showarticletitle{Interdepartmental integration: a definition with
  implications for product development performance}.
\newblock \bibinfo{journal}{\emph{Journal of product innovation management}}
  \bibinfo{volume}{13}, \bibinfo{number}{2} (\bibinfo{year}{1996}),
  \bibinfo{pages}{137--151}.
\newblock


\bibitem[Kaur et~al\mbox{.}(2020)]%
        {Kaur2020InterpretingIU}
\bibfield{author}{\bibinfo{person}{Harmanpreet Kaur}, \bibinfo{person}{Harsha
  Nori}, \bibinfo{person}{Samuel Jenkins}, \bibinfo{person}{Rich Caruana},
  \bibinfo{person}{Hanna~M. Wallach}, {and} \bibinfo{person}{Jennifer~Wortman
  Vaughan}.} \bibinfo{year}{2020}\natexlab{}.
\newblock \showarticletitle{Interpreting Interpretability: Understanding Data
  Scientists' Use of Interpretability Tools for Machine Learning}.
\newblock \bibinfo{journal}{\emph{Proceedings of the 2020 CHI Conference on
  Human Factors in Computing Systems}} (\bibinfo{year}{2020}).
\newblock


\bibitem[Kleinberg et~al\mbox{.}(2016)]%
        {kleinberg2016inherent}
\bibfield{author}{\bibinfo{person}{Jon Kleinberg}, \bibinfo{person}{Sendhil
  Mullainathan}, {and} \bibinfo{person}{Manish Raghavan}.}
  \bibinfo{year}{2016}\natexlab{}.
\newblock \showarticletitle{Inherent trade-offs in the fair determination of
  risk scores}.
\newblock \bibinfo{journal}{\emph{arXiv preprint arXiv:1609.05807}}
  (\bibinfo{year}{2016}).
\newblock


\bibitem[Kross and Guo(2021)]%
        {kross2021orienting}
\bibfield{author}{\bibinfo{person}{Sean Kross} {and} \bibinfo{person}{Philip
  Guo}.} \bibinfo{year}{2021}\natexlab{}.
\newblock \showarticletitle{Orienting, framing, bridging, magic, and
  counseling: How data scientists navigate the outer loop of client
  collaborations in industry and academia}.
\newblock \bibinfo{journal}{\emph{Proceedings of the ACM on Human-Computer
  Interaction}} \bibinfo{volume}{5}, \bibinfo{number}{CSCW2}
  (\bibinfo{year}{2021}), \bibinfo{pages}{1--28}.
\newblock


\bibitem[Lam et~al\mbox{.}(2023)]%
        {lam2023model}
\bibfield{author}{\bibinfo{person}{Michelle~S Lam}, \bibinfo{person}{Zixian
  Ma}, \bibinfo{person}{Anne Li}, \bibinfo{person}{Izequiel Freitas},
  \bibinfo{person}{Dakuo Wang}, \bibinfo{person}{James~A Landay}, {and}
  \bibinfo{person}{Michael~S Bernstein}.} \bibinfo{year}{2023}\natexlab{}.
\newblock \showarticletitle{Model Sketching: Centering Concepts in Early-Stage
  Machine Learning Model Design}. In \bibinfo{booktitle}{\emph{Proceedings of
  the 2023 CHI Conference on Human Factors in Computing Systems}}.
  \bibinfo{pages}{1--24}.
\newblock


\bibitem[Lee et~al\mbox{.}(2022)]%
        {lee2022hci}
\bibfield{author}{\bibinfo{person}{Jung-Joo Lee}, \bibinfo{person}{Christine
  Ee~Ling Yap}, {and} \bibinfo{person}{Virpi Roto}.}
  \bibinfo{year}{2022}\natexlab{}.
\newblock \showarticletitle{How HCI Adopts Service Design: Unpacking current
  perceptions and scopes of service design in HCI and identifying future
  opportunities}. In \bibinfo{booktitle}{\emph{CHI Conference on Human Factors
  in Computing Systems}}. \bibinfo{pages}{1--14}.
\newblock


\bibitem[Lee et~al\mbox{.}(2021)]%
        {lee2021formalising}
\bibfield{author}{\bibinfo{person}{Michelle Seng~Ah Lee},
  \bibinfo{person}{Luciano Floridi}, {and} \bibinfo{person}{Jatinder Singh}.}
  \bibinfo{year}{2021}\natexlab{}.
\newblock \showarticletitle{Formalising trade-offs beyond algorithmic fairness:
  lessons from ethical philosophy and welfare economics}.
\newblock \bibinfo{journal}{\emph{AI and Ethics}} \bibinfo{volume}{1},
  \bibinfo{number}{4} (\bibinfo{year}{2021}), \bibinfo{pages}{529--544}.
\newblock


\bibitem[Lee and Singh(2021)]%
        {lee2021landscape}
\bibfield{author}{\bibinfo{person}{Michelle Seng~Ah Lee} {and}
  \bibinfo{person}{Jat Singh}.} \bibinfo{year}{2021}\natexlab{}.
\newblock \showarticletitle{The landscape and gaps in open source fairness
  toolkits}. In \bibinfo{booktitle}{\emph{Proceedings of the 2021 CHI
  conference on human factors in computing systems}}. \bibinfo{pages}{1--13}.
\newblock


\bibitem[Lewicki et~al\mbox{.}(2023)]%
        {lewicki2023out}
\bibfield{author}{\bibinfo{person}{Kornel Lewicki}, \bibinfo{person}{Michelle
  Seng~Ah Lee}, \bibinfo{person}{Jennifer Cobbe}, {and}
  \bibinfo{person}{Jatinder Singh}.} \bibinfo{year}{2023}\natexlab{}.
\newblock \showarticletitle{Out of Context: Investigating the Bias and Fairness
  Concerns of “Artificial Intelligence as a Service”}. In
  \bibinfo{booktitle}{\emph{Proceedings of the 2023 CHI Conference on Human
  Factors in Computing Systems}}. \bibinfo{pages}{1--17}.
\newblock


\bibitem[Li et~al\mbox{.}(2018)]%
        {li2018coconut}
\bibfield{author}{\bibinfo{person}{Tianshi Li}, \bibinfo{person}{Yuvraj
  Agarwal}, {and} \bibinfo{person}{Jason~I Hong}.}
  \bibinfo{year}{2018}\natexlab{}.
\newblock \showarticletitle{Coconut: An IDE plugin for developing
  privacy-friendly apps}.
\newblock \bibinfo{journal}{\emph{Proceedings of the ACM on Interactive,
  Mobile, Wearable and Ubiquitous Technologies}} \bibinfo{volume}{2},
  \bibinfo{number}{4} (\bibinfo{year}{2018}), \bibinfo{pages}{1--35}.
\newblock


\bibitem[Liao et~al\mbox{.}(2023)]%
        {liao2023designerly}
\bibfield{author}{\bibinfo{person}{Q~Vera Liao}, \bibinfo{person}{Hariharan
  Subramonyam}, \bibinfo{person}{Jennifer Wang}, {and}
  \bibinfo{person}{Jennifer Wortman~Vaughan}.} \bibinfo{year}{2023}\natexlab{}.
\newblock \showarticletitle{Designerly Understanding: Information Needs for
  Model Transparency to Support Design Ideation for AI-Powered User
  Experience}. In \bibinfo{booktitle}{\emph{Proceedings of the 2023 CHI
  Conference on Human Factors in Computing Systems}}. \bibinfo{pages}{1--21}.
\newblock


\bibitem[Liskov(1987)]%
        {liskov1987keynote}
\bibfield{author}{\bibinfo{person}{Barbara Liskov}.}
  \bibinfo{year}{1987}\natexlab{}.
\newblock \showarticletitle{Keynote address-data abstraction and hierarchy}. In
  \bibinfo{booktitle}{\emph{Addendum to the proceedings on Object-oriented
  programming systems, languages and applications (Addendum)}}.
  \bibinfo{pages}{17--34}.
\newblock


\bibitem[Madaio et~al\mbox{.}(2022)]%
        {madaio2022assessing}
\bibfield{author}{\bibinfo{person}{Michael Madaio}, \bibinfo{person}{Lisa
  Egede}, \bibinfo{person}{Hariharan Subramonyam}, \bibinfo{person}{Jennifer
  Wortman~Vaughan}, {and} \bibinfo{person}{Hanna Wallach}.}
  \bibinfo{year}{2022}\natexlab{}.
\newblock \showarticletitle{Assessing the Fairness of AI Systems: AI
  Practitioners' Processes, Challenges, and Needs for Support}.
\newblock \bibinfo{journal}{\emph{Proceedings of the ACM on Human-Computer
  Interaction}} \bibinfo{volume}{6}, \bibinfo{number}{CSCW1}
  (\bibinfo{year}{2022}), \bibinfo{pages}{1--26}.
\newblock


\bibitem[Madaio et~al\mbox{.}(2020a)]%
        {madaio2020need}
\bibfield{author}{\bibinfo{person}{Michael Madaio}, \bibinfo{person}{Luke
  Stark}, \bibinfo{person}{Wortman~Vaughan Jennifer}, {and}
  \bibinfo{person}{Hanna Wallach}.} \bibinfo{year}{2020}\natexlab{a}.
\newblock \showarticletitle{Need for Organizational Performance Metrics to
  Support Fairness in AI}.
\newblock \bibinfo{journal}{\emph{Fair and Responsible AI Workshop at the 2020
  CHI Conference on Human Factors in Computing Systems}}
  (\bibinfo{year}{2020}), \bibinfo{pages}{1--14}.
\newblock


\bibitem[Madaio et~al\mbox{.}(2020b)]%
        {madaio2020co}
\bibfield{author}{\bibinfo{person}{Michael~A Madaio}, \bibinfo{person}{Luke
  Stark}, \bibinfo{person}{Jennifer Wortman~Vaughan}, {and}
  \bibinfo{person}{Hanna Wallach}.} \bibinfo{year}{2020}\natexlab{b}.
\newblock \showarticletitle{Co-designing checklists to understand
  organizational challenges and opportunities around fairness in ai}. In
  \bibinfo{booktitle}{\emph{Proceedings of the 2020 CHI Conference on Human
  Factors in Computing Systems}}. \bibinfo{pages}{1--14}.
\newblock


\bibitem[Mao et~al\mbox{.}(2019)]%
        {mao2019data}
\bibfield{author}{\bibinfo{person}{Yaoli Mao}, \bibinfo{person}{Dakuo Wang},
  \bibinfo{person}{Michael Muller}, \bibinfo{person}{Kush~R Varshney},
  \bibinfo{person}{Ioana Baldini}, \bibinfo{person}{Casey Dugan}, {and}
  \bibinfo{person}{Aleksandra Mojsilovi{\'c}}.}
  \bibinfo{year}{2019}\natexlab{}.
\newblock \showarticletitle{How data scientistswork together with domain
  experts in scientific collaborations: To find the right answer or to ask the
  right question?}
\newblock \bibinfo{journal}{\emph{Proceedings of the ACM on Human-Computer
  Interaction}} \bibinfo{volume}{3}, \bibinfo{number}{GROUP}
  (\bibinfo{year}{2019}), \bibinfo{pages}{1--23}.
\newblock


\bibitem[Metcalf et~al\mbox{.}(2019)]%
        {metcalf2019owning}
\bibfield{author}{\bibinfo{person}{Jacob Metcalf}, \bibinfo{person}{Emanuel
  Moss}, {et~al\mbox{.}}} \bibinfo{year}{2019}\natexlab{}.
\newblock \showarticletitle{Owning ethics: Corporate logics, silicon valley,
  and the institutionalization of ethics}.
\newblock \bibinfo{journal}{\emph{Social Research: An International Quarterly}}
  \bibinfo{volume}{86}, \bibinfo{number}{2} (\bibinfo{year}{2019}),
  \bibinfo{pages}{449--476}.
\newblock


\bibitem[Moore et~al\mbox{.}(2023)]%
        {moore2023failurenotes}
\bibfield{author}{\bibinfo{person}{Steven Moore}, \bibinfo{person}{Q~Vera
  Liao}, {and} \bibinfo{person}{Hariharan Subramonyam}.}
  \bibinfo{year}{2023}\natexlab{}.
\newblock \showarticletitle{fAIlureNotes: Supporting Designers in Understanding
  the Limits of AI Models for Computer Vision Tasks}. In
  \bibinfo{booktitle}{\emph{Proceedings of the 2023 CHI Conference on Human
  Factors in Computing Systems}}. \bibinfo{pages}{1--19}.
\newblock


\bibitem[Muller et~al\mbox{.}(2019)]%
        {muller2019data}
\bibfield{author}{\bibinfo{person}{Michael Muller}, \bibinfo{person}{Ingrid
  Lange}, \bibinfo{person}{Dakuo Wang}, \bibinfo{person}{David Piorkowski},
  \bibinfo{person}{Jason Tsay}, \bibinfo{person}{Q~Vera Liao},
  \bibinfo{person}{Casey Dugan}, {and} \bibinfo{person}{Thomas Erickson}.}
  \bibinfo{year}{2019}\natexlab{}.
\newblock \showarticletitle{How data science workers work with data: Discovery,
  capture, curation, design, creation}. In
  \bibinfo{booktitle}{\emph{Proceedings of the 2019 CHI conference on human
  factors in computing systems}}. \bibinfo{pages}{1--15}.
\newblock


\bibitem[Nafus and Sherman(2014)]%
        {Nafus2014ThisOD}
\bibfield{author}{\bibinfo{person}{Dawn Nafus} {and} \bibinfo{person}{Jamie
  Sherman}.} \bibinfo{year}{2014}\natexlab{}.
\newblock \showarticletitle{This One Does Not Go Up to 11: The Quantified Self
  Movement as an Alternative Big Data Practice}.
\newblock


\bibitem[Nahar et~al\mbox{.}(2021)]%
        {nahar2021collaboration}
\bibfield{author}{\bibinfo{person}{Nadia Nahar}, \bibinfo{person}{Shurui Zhou},
  \bibinfo{person}{Grace Lewis}, {and} \bibinfo{person}{Christian
  K{\"a}stner}.} \bibinfo{year}{2021}\natexlab{}.
\newblock \showarticletitle{Collaboration Challenges in Building ML-Enabled
  Systems: Communication, Documentation, Engineering, and Process}.
\newblock \bibinfo{journal}{\emph{arXiv preprint arXiv:2110.10234}}
  (\bibinfo{year}{2021}).
\newblock


\bibitem[Obermeyer et~al\mbox{.}(2019)]%
        {obermeyer2019dissecting}
\bibfield{author}{\bibinfo{person}{Ziad Obermeyer}, \bibinfo{person}{Brian
  Powers}, \bibinfo{person}{Christine Vogeli}, {and} \bibinfo{person}{Sendhil
  Mullainathan}.} \bibinfo{year}{2019}\natexlab{}.
\newblock \showarticletitle{Dissecting racial bias in an algorithm used to
  manage the health of populations}.
\newblock \bibinfo{journal}{\emph{Science}} \bibinfo{volume}{366},
  \bibinfo{number}{6464} (\bibinfo{year}{2019}), \bibinfo{pages}{447--453}.
\newblock


\bibitem[Park et~al\mbox{.}(2021)]%
        {park2021facilitating}
\bibfield{author}{\bibinfo{person}{Soya Park}, \bibinfo{person}{April~Yi Wang},
  \bibinfo{person}{Ban Kawas}, \bibinfo{person}{Q~Vera Liao},
  \bibinfo{person}{David Piorkowski}, {and} \bibinfo{person}{Marina
  Danilevsky}.} \bibinfo{year}{2021}\natexlab{}.
\newblock \showarticletitle{Facilitating knowledge sharing from domain experts
  to data scientists for building nlp models}. In
  \bibinfo{booktitle}{\emph{26th International Conference on Intelligent User
  Interfaces}}. \bibinfo{pages}{585--596}.
\newblock


\bibitem[Passi and Barocas(2019)]%
        {passi2019problem}
\bibfield{author}{\bibinfo{person}{Samir Passi} {and} \bibinfo{person}{Solon
  Barocas}.} \bibinfo{year}{2019}\natexlab{}.
\newblock \showarticletitle{Problem formulation and fairness}. In
  \bibinfo{booktitle}{\emph{Proceedings of the Conference on Fairness,
  Accountability, and Transparency}}. \bibinfo{pages}{39--48}.
\newblock


\bibitem[Passi and Jackson(2018)]%
        {passi2018trust}
\bibfield{author}{\bibinfo{person}{Samir Passi} {and} \bibinfo{person}{Steven~J
  Jackson}.} \bibinfo{year}{2018}\natexlab{}.
\newblock \showarticletitle{Trust in data science: Collaboration, translation,
  and accountability in corporate data science projects}.
\newblock \bibinfo{journal}{\emph{Proceedings of the ACM on Human-Computer
  Interaction}} \bibinfo{volume}{2}, \bibinfo{number}{CSCW}
  (\bibinfo{year}{2018}), \bibinfo{pages}{1--28}.
\newblock


\bibitem[Pavelin et~al\mbox{.}(2014)]%
        {pavelin2014ten}
\bibfield{author}{\bibinfo{person}{Katrina Pavelin}, \bibinfo{person}{Sangya
  Pundir}, {and} \bibinfo{person}{Jennifer~A Cham}.}
  \bibinfo{year}{2014}\natexlab{}.
\newblock \showarticletitle{Ten simple rules for running interactive
  workshops}.
\newblock \bibinfo{journal}{\emph{PLoS computational biology}}
  \bibinfo{volume}{10}, \bibinfo{number}{2} (\bibinfo{year}{2014}),
  \bibinfo{pages}{e1003485}.
\newblock


\bibitem[Piorkowski et~al\mbox{.}(2021)]%
        {piorkowski2021ai}
\bibfield{author}{\bibinfo{person}{David Piorkowski}, \bibinfo{person}{Soya
  Park}, \bibinfo{person}{April~Yi Wang}, \bibinfo{person}{Dakuo Wang},
  \bibinfo{person}{Michael Muller}, {and} \bibinfo{person}{Felix Portnoy}.}
  \bibinfo{year}{2021}\natexlab{}.
\newblock \showarticletitle{How ai developers overcome communication challenges
  in a multidisciplinary team: A case study}.
\newblock \bibinfo{journal}{\emph{Proceedings of the ACM on Human-Computer
  Interaction}} \bibinfo{volume}{5}, \bibinfo{number}{CSCW1}
  (\bibinfo{year}{2021}), \bibinfo{pages}{1--25}.
\newblock


\bibitem[Raji et~al\mbox{.}(2020)]%
        {raji2020saving}
\bibfield{author}{\bibinfo{person}{Inioluwa~Deborah Raji},
  \bibinfo{person}{Timnit Gebru}, \bibinfo{person}{Margaret Mitchell},
  \bibinfo{person}{Joy Buolamwini}, \bibinfo{person}{Joonseok Lee}, {and}
  \bibinfo{person}{Emily Denton}.} \bibinfo{year}{2020}\natexlab{}.
\newblock \showarticletitle{Saving face: Investigating the ethical concerns of
  facial recognition auditing}. In \bibinfo{booktitle}{\emph{Proceedings of the
  AAAI/ACM Conference on AI, Ethics, and Society}}. \bibinfo{pages}{145--151}.
\newblock


\bibitem[Rakova et~al\mbox{.}(2021)]%
        {rakova2021responsible}
\bibfield{author}{\bibinfo{person}{Bogdana Rakova}, \bibinfo{person}{Jingying
  Yang}, \bibinfo{person}{Henriette Cramer}, {and} \bibinfo{person}{Rumman
  Chowdhury}.} \bibinfo{year}{2021}\natexlab{}.
\newblock \showarticletitle{Where responsible AI meets reality: Practitioner
  perspectives on enablers for shifting organizational practices}.
\newblock \bibinfo{journal}{\emph{Proceedings of the ACM on Human-Computer
  Interaction}} \bibinfo{volume}{5}, \bibinfo{number}{CSCW1}
  (\bibinfo{year}{2021}), \bibinfo{pages}{1--23}.
\newblock


\bibitem[Richardson et~al\mbox{.}(2021)]%
        {richardson2021towards}
\bibfield{author}{\bibinfo{person}{Brianna Richardson}, \bibinfo{person}{Jean
  Garcia-Gathright}, \bibinfo{person}{Samuel~F Way}, \bibinfo{person}{Jennifer
  Thom}, {and} \bibinfo{person}{Henriette Cramer}.}
  \bibinfo{year}{2021}\natexlab{}.
\newblock \showarticletitle{Towards Fairness in Practice: A
  Practitioner-Oriented Rubric for Evaluating Fair ML Toolkits}. In
  \bibinfo{booktitle}{\emph{Proceedings of the 2021 CHI Conference on Human
  Factors in Computing Systems}}. \bibinfo{pages}{1--13}.
\newblock


\bibitem[Robinson(2010)]%
        {robinson2010save}
\bibfield{author}{\bibinfo{person}{Thomas~N Robinson}.}
  \bibinfo{year}{2010}\natexlab{}.
\newblock \showarticletitle{Save the world, prevent obesity: piggybacking on
  existing social and ideological movements}.
\newblock \bibinfo{journal}{\emph{Obesity}} \bibinfo{volume}{18},
  \bibinfo{number}{n1s} (\bibinfo{year}{2010}), \bibinfo{pages}{S17}.
\newblock


\bibitem[Rostamzadeh et~al\mbox{.}(2022)]%
        {rostamzadeh2022healthsheet}
\bibfield{author}{\bibinfo{person}{Negar Rostamzadeh}, \bibinfo{person}{Diana
  Mincu}, \bibinfo{person}{Subhrajit Roy}, \bibinfo{person}{Andrew Smart},
  \bibinfo{person}{Lauren Wilcox}, \bibinfo{person}{Mahima Pushkarna},
  \bibinfo{person}{Jessica Schrouff}, \bibinfo{person}{Razvan Amironesei},
  \bibinfo{person}{Nyalleng Moorosi}, {and} \bibinfo{person}{Katherine
  Heller}.} \bibinfo{year}{2022}\natexlab{}.
\newblock \showarticletitle{Healthsheet: development of a transparency artifact
  for health datasets}. In \bibinfo{booktitle}{\emph{2022 ACM Conference on
  Fairness, Accountability, and Transparency}}. \bibinfo{pages}{1943--1961}.
\newblock


\bibitem[Scott and Davis(2015)]%
        {scott2015organizations}
\bibfield{author}{\bibinfo{person}{W~Richard Scott} {and}
  \bibinfo{person}{Gerald~F Davis}.} \bibinfo{year}{2015}\natexlab{}.
\newblock \bibinfo{booktitle}{\emph{Organizations and organizing: Rational,
  natural and open systems perspectives}}.
\newblock \bibinfo{publisher}{Routledge}.
\newblock


\bibitem[Selbst et~al\mbox{.}(2019)]%
        {selbst2019fairness}
\bibfield{author}{\bibinfo{person}{Andrew~D Selbst}, \bibinfo{person}{Danah
  Boyd}, \bibinfo{person}{Sorelle~A Friedler}, \bibinfo{person}{Suresh
  Venkatasubramanian}, {and} \bibinfo{person}{Janet Vertesi}.}
  \bibinfo{year}{2019}\natexlab{}.
\newblock \showarticletitle{Fairness and abstraction in sociotechnical
  systems}. In \bibinfo{booktitle}{\emph{Proceedings of the conference on
  fairness, accountability, and transparency}}. \bibinfo{pages}{59--68}.
\newblock


\bibitem[Shen et~al\mbox{.}(2021a)]%
        {shen2021value}
\bibfield{author}{\bibinfo{person}{Hong Shen}, \bibinfo{person}{Wesley~H Deng},
  \bibinfo{person}{Aditi Chattopadhyay}, \bibinfo{person}{Zhiwei~Steven Wu},
  \bibinfo{person}{Xu Wang}, {and} \bibinfo{person}{Haiyi Zhu}.}
  \bibinfo{year}{2021}\natexlab{a}.
\newblock \showarticletitle{Value cards: An educational toolkit for teaching
  social impacts of machine learning through deliberation}. In
  \bibinfo{booktitle}{\emph{Proceedings of the 2021 ACM conference on fairness,
  accountability, and transparency}}. \bibinfo{pages}{850--861}.
\newblock


\bibitem[Shen et~al\mbox{.}(2021b)]%
        {shen2021everyday}
\bibfield{author}{\bibinfo{person}{Hong Shen}, \bibinfo{person}{Alicia DeVos},
  \bibinfo{person}{Motahhare Eslami}, {and} \bibinfo{person}{Kenneth
  Holstein}.} \bibinfo{year}{2021}\natexlab{b}.
\newblock \showarticletitle{Everyday algorithm auditing: Understanding the
  power of everyday users in surfacing harmful algorithmic behaviors}.
\newblock \bibinfo{journal}{\emph{Proceedings of the ACM on Human-Computer
  Interaction}} \bibinfo{volume}{5}, \bibinfo{number}{CSCW2}
  (\bibinfo{year}{2021}), \bibinfo{pages}{1--29}.
\newblock


\bibitem[Shneiderman(2020)]%
        {Shneiderman2020HumanCenteredAI}
\bibfield{author}{\bibinfo{person}{Ben Shneiderman}.}
  \bibinfo{year}{2020}\natexlab{}.
\newblock \showarticletitle{Human-Centered Artificial Intelligence: Reliable,
  Safe \& Trustworthy}.
\newblock \bibinfo{journal}{\emph{International Journal of Human–Computer
  Interaction}}  \bibinfo{volume}{36} (\bibinfo{year}{2020}),
  \bibinfo{pages}{495 -- 504}.
\newblock


\bibitem[Small and Calarco(2022)]%
        {small2022qualitative}
\bibfield{author}{\bibinfo{person}{Mario~Luis Small} {and}
  \bibinfo{person}{Jessica~McCrory Calarco}.} \bibinfo{year}{2022}\natexlab{}.
\newblock \bibinfo{booktitle}{\emph{Qualitative literacy: A guide to evaluating
  ethnographic and interview research}}.
\newblock \bibinfo{publisher}{Univ of California Press}.
\newblock


\bibitem[Smith et~al\mbox{.}(2022)]%
        {smith2022real}
\bibfield{author}{\bibinfo{person}{Jessie~J Smith}, \bibinfo{person}{Saleema
  Amershi}, \bibinfo{person}{Solon Barocas}, \bibinfo{person}{Hanna Wallach},
  {and} \bibinfo{person}{Jennifer Wortman~Vaughan}.}
  \bibinfo{year}{2022}\natexlab{}.
\newblock \showarticletitle{REAL ML: Recognizing, Exploring, and Articulating
  Limitations of Machine Learning Research}. In \bibinfo{booktitle}{\emph{2022
  ACM Conference on Fairness, Accountability, and Transparency}}.
  \bibinfo{pages}{587--597}.
\newblock


\bibitem[Song and Parry(1997)]%
        {song1997cross}
\bibfield{author}{\bibinfo{person}{X~Michael Song} {and}
  \bibinfo{person}{Mark~E Parry}.} \bibinfo{year}{1997}\natexlab{}.
\newblock \showarticletitle{A cross-national comparative study of new product
  development processes: Japan and the United States}.
\newblock \bibinfo{journal}{\emph{Journal of marketing}} \bibinfo{volume}{61},
  \bibinfo{number}{2} (\bibinfo{year}{1997}), \bibinfo{pages}{1--18}.
\newblock


\bibitem[Star and Strauss(1999)]%
        {star1999layers}
\bibfield{author}{\bibinfo{person}{Susan~Leigh Star} {and}
  \bibinfo{person}{Anselm Strauss}.} \bibinfo{year}{1999}\natexlab{}.
\newblock \showarticletitle{Layers of silence, arenas of voice: The ecology of
  visible and invisible work}.
\newblock \bibinfo{journal}{\emph{Computer supported cooperative work}}
  \bibinfo{volume}{8}, \bibinfo{number}{1-2} (\bibinfo{year}{1999}),
  \bibinfo{pages}{9--30}.
\newblock


\bibitem[Subramonyam et~al\mbox{.}(2022)]%
        {subramonyam2022solving}
\bibfield{author}{\bibinfo{person}{Hariharan Subramonyam},
  \bibinfo{person}{Jane Im}, \bibinfo{person}{Colleen Seifert}, {and}
  \bibinfo{person}{Eytan Adar}.} \bibinfo{year}{2022}\natexlab{}.
\newblock \showarticletitle{Solving Separation-of-Concerns Problems in
  Collaborative Design of Human-AI Systems through Leaky Abstractions}. In
  \bibinfo{booktitle}{\emph{CHI Conference on Human Factors in Computing
  Systems}}. \bibinfo{pages}{1--21}.
\newblock


\bibitem[Subramonyam et~al\mbox{.}(2021)]%
        {subramonyam2021towards}
\bibfield{author}{\bibinfo{person}{Hariharan Subramonyam},
  \bibinfo{person}{Colleen Seifert}, {and} \bibinfo{person}{Eytan Adar}.}
  \bibinfo{year}{2021}\natexlab{}.
\newblock \showarticletitle{Towards a process model for co-creating AI
  experiences}. In \bibinfo{booktitle}{\emph{Designing Interactive Systems
  Conference 2021}}. \bibinfo{pages}{1529--1543}.
\newblock


\bibitem[Tahaei et~al\mbox{.}(2021)]%
        {tahaei2021privacy}
\bibfield{author}{\bibinfo{person}{Mohammad Tahaei}, \bibinfo{person}{Alisa
  Frik}, {and} \bibinfo{person}{Kami Vaniea}.} \bibinfo{year}{2021}\natexlab{}.
\newblock \showarticletitle{Privacy champions in software teams: understanding
  their motivations, strategies, and challenges}. In
  \bibinfo{booktitle}{\emph{Proceedings of the 2021 CHI Conference on Human
  Factors in Computing Systems}}. \bibinfo{pages}{1--15}.
\newblock


\bibitem[Varshney(2019)]%
        {varshney2019trustworthy}
\bibfield{author}{\bibinfo{person}{Kush~R Varshney}.}
  \bibinfo{year}{2019}\natexlab{}.
\newblock \showarticletitle{Trustworthy machine learning and artificial
  intelligence}.
\newblock \bibinfo{journal}{\emph{XRDS: Crossroads, The ACM Magazine for
  Students}} \bibinfo{volume}{25}, \bibinfo{number}{3} (\bibinfo{year}{2019}),
  \bibinfo{pages}{26--29}.
\newblock


\bibitem[Waldman(2021)]%
        {waldman2021industry}
\bibfield{author}{\bibinfo{person}{Ari~Ezra Waldman}.}
  \bibinfo{year}{2021}\natexlab{}.
\newblock \bibinfo{booktitle}{\emph{Industry unbound: The inside story of
  privacy, data, and corporate power}}.
\newblock \bibinfo{publisher}{Cambridge University Press}.
\newblock


\bibitem[Wang et~al\mbox{.}(2022)]%
        {wang2022measuring}
\bibfield{author}{\bibinfo{person}{Angelina Wang}, \bibinfo{person}{Solon
  Barocas}, \bibinfo{person}{Kristen Laird}, {and} \bibinfo{person}{Hanna
  Wallach}.} \bibinfo{year}{2022}\natexlab{}.
\newblock \showarticletitle{Measuring representational harms in image
  captioning}. In \bibinfo{booktitle}{\emph{2022 ACM Conference on Fairness,
  Accountability, and Transparency}}. \bibinfo{pages}{324--335}.
\newblock


\bibitem[Wang et~al\mbox{.}(2023)]%
        {wang2023designing}
\bibfield{author}{\bibinfo{person}{Qiaosi Wang}, \bibinfo{person}{Michael~Adam
  Madaio}, \bibinfo{person}{Shivani Kapania}, \bibinfo{person}{Shaun Kane},
  \bibinfo{person}{Michael Terry}, \bibinfo{person}{Lauren Wilcox},
  {et~al\mbox{.}}} \bibinfo{year}{2023}\natexlab{}.
\newblock \showarticletitle{Designing Responsible AI: Adaptations of UX
  Practice to Meet Responsible AI Challenges}.
\newblock  (\bibinfo{year}{2023}).
\newblock


\bibitem[Wong(2021)]%
        {wong2021tactics}
\bibfield{author}{\bibinfo{person}{Richmond~Y Wong}.}
  \bibinfo{year}{2021}\natexlab{}.
\newblock \showarticletitle{Tactics of Soft Resistance in User Experience
  Professionals' Values Work}.
\newblock \bibinfo{journal}{\emph{Proceedings of the ACM on Human-Computer
  Interaction}} \bibinfo{volume}{5}, \bibinfo{number}{CSCW2}
  (\bibinfo{year}{2021}), \bibinfo{pages}{1--28}.
\newblock


\bibitem[Wong et~al\mbox{.}(2022)]%
        {wong2022seeing}
\bibfield{author}{\bibinfo{person}{Richmond~Y Wong}, \bibinfo{person}{Michael~A
  Madaio}, {and} \bibinfo{person}{Nick Merrill}.}
  \bibinfo{year}{2022}\natexlab{}.
\newblock \showarticletitle{Seeing Like a Toolkit: How Toolkits Envision the
  Work of AI Ethics}.
\newblock \bibinfo{journal}{\emph{arXiv preprint arXiv:2202.08792}}
  (\bibinfo{year}{2022}).
\newblock


\bibitem[Wong and Nguyen(2021)]%
        {wong2021timelines}
\bibfield{author}{\bibinfo{person}{Richmond~Y Wong} {and}
  \bibinfo{person}{Tonya Nguyen}.} \bibinfo{year}{2021}\natexlab{}.
\newblock \showarticletitle{Timelines: A world-building activity for values
  advocacy}. In \bibinfo{booktitle}{\emph{Proceedings of the 2021 CHI
  Conference on Human Factors in Computing Systems}}. \bibinfo{pages}{1--15}.
\newblock


\bibitem[Yang et~al\mbox{.}(2020)]%
        {yang2020re}
\bibfield{author}{\bibinfo{person}{Qian Yang}, \bibinfo{person}{Aaron
  Steinfeld}, \bibinfo{person}{Carolyn Ros{\'e}}, {and} \bibinfo{person}{John
  Zimmerman}.} \bibinfo{year}{2020}\natexlab{}.
\newblock \showarticletitle{Re-examining whether, why, and how human-AI
  interaction is uniquely difficult to design}. In
  \bibinfo{booktitle}{\emph{Proceedings of the 2020 chi conference on human
  factors in computing systems}}. \bibinfo{pages}{1--13}.
\newblock


\bibitem[Yildirim et~al\mbox{.}(2022)]%
        {yildirim2022experienced}
\bibfield{author}{\bibinfo{person}{Nur Yildirim}, \bibinfo{person}{Alex Kass},
  \bibinfo{person}{Teresa Tung}, \bibinfo{person}{Connor Upton},
  \bibinfo{person}{Donnacha Costello}, \bibinfo{person}{Robert Giusti},
  \bibinfo{person}{Sinem Lacin}, \bibinfo{person}{Sara Lovic},
  \bibinfo{person}{James~M O'Neill}, \bibinfo{person}{Rudi~O'Reilly Meehan},
  {et~al\mbox{.}}} \bibinfo{year}{2022}\natexlab{}.
\newblock \showarticletitle{How Experienced Designers of Enterprise
  Applications Engage AI as a Design Material}. In
  \bibinfo{booktitle}{\emph{Proceedings of the 2022 CHI Conference on Human
  Factors in Computing Systems}}. \bibinfo{pages}{1--13}.
\newblock


\bibitem[Yildirim et~al\mbox{.}(2023)]%
        {yildirim2023investigating}
\bibfield{author}{\bibinfo{person}{Nur Yildirim}, \bibinfo{person}{Mahima
  Pushkarna}, \bibinfo{person}{Nitesh Goyal}, \bibinfo{person}{Martin
  Wattenberg}, {and} \bibinfo{person}{Fernanda Vi{\'e}gas}.}
  \bibinfo{year}{2023}\natexlab{}.
\newblock \showarticletitle{Investigating How Practitioners Use Human-AI
  Guidelines: A Case Study on the People+ AI Guidebook}. In
  \bibinfo{booktitle}{\emph{Proceedings of the 2023 CHI Conference on Human
  Factors in Computing Systems}}. \bibinfo{pages}{1--13}.
\newblock


\bibitem[Yurrita et~al\mbox{.}(2023)]%
        {yurrita2023disentangling}
\bibfield{author}{\bibinfo{person}{Mireia Yurrita}, \bibinfo{person}{Tim
  Draws}, \bibinfo{person}{Agathe Balayn}, \bibinfo{person}{Dave Murray-Rust},
  \bibinfo{person}{Nava Tintarev}, {and} \bibinfo{person}{Alessandro Bozzon}.}
  \bibinfo{year}{2023}\natexlab{}.
\newblock \showarticletitle{Disentangling Fairness Perceptions in Algorithmic
  Decision-Making: the Effects of Explanations, Human Oversight, and
  Contestability}. In \bibinfo{booktitle}{\emph{Proceedings of the 2023 CHI
  Conference on Human Factors in Computing Systems}}. \bibinfo{pages}{1--21}.
\newblock


\bibitem[Zhang et~al\mbox{.}(2020)]%
        {zhang2020data}
\bibfield{author}{\bibinfo{person}{Amy~X Zhang}, \bibinfo{person}{Michael
  Muller}, {and} \bibinfo{person}{Dakuo Wang}.}
  \bibinfo{year}{2020}\natexlab{}.
\newblock \showarticletitle{How do data science workers collaborate? roles,
  workflows, and tools}.
\newblock \bibinfo{journal}{\emph{Proceedings of the ACM on Human-Computer
  Interaction}} \bibinfo{volume}{4}, \bibinfo{number}{CSCW1}
  (\bibinfo{year}{2020}), \bibinfo{pages}{1--23}.
\newblock


\end{thebibliography}

%\appendix

%\section{Appendix}

%%
%% If your work has an appendix, this is the place to put it.

\end{document}